\documentclass[aps,prb,twocolumn, superscriptaddress]
{revtex4-1}

\usepackage{amsmath} 
\usepackage{amsthm} 
\usepackage{graphicx} 
\begin{document}

\title{Field-dependent nonlinear surface resistance and its optimization by surface nano-structuring in superconductors}

\author{Takayuki Kubo}
\affiliation{KEK (High Energy Accelerator Research Organization), Tsukuba, Ibaraki 305-0801, Japan.}
\affiliation{SOKENDAI (the Graduate University for Advanced Studies), Hayama, Kanagawa 240-0115, Japan.}
\affiliation{Department of Physics, and Center for Accelerator Science, Old Dominion University, Norfolk, Virginia 23529, USA.}

\author{Alex Gurevich}
\affiliation{Department of Physics, and Center for Accelerator Science, Old Dominion University, Norfolk, Virginia 23529, USA.}



\begin{abstract}
We propose a theory of nonlinear surface resistance of a dirty superconductor in a strong radio-frequency (RF) field, taking into account magnetic and nonmagnetic impurities, finite quasiparticle lifetimes, and a thin proximity-coupled normal layer characteristic of the oxide surface of many materials. The Usadel equations were solved to obtain the quasiparticle density of states (DOS) and the low-frequency surface resistance $R_s$ as functions of the RF field amplitude $H_0$. 
It is shown that the interplay of the broadening of the DOS peaks and a decrease of a quasiparticle gap caused by the RF currents produces a minimum in $R_s(H_0)$ and an extended rise of the quality factor $Q(H_0)$ with the RF field.
Paramagnetic impurities shift the minimum in $R_s(H_0)$ to lower fields and can reduce $R_s(H_0)$ in a wide range of $H_0$. 
Subgap states in the DOS can give rise to a residual surface resistance while reducing $R_s$ at higher temperatures. 
A proximity-coupled normal layer at the surface can shift the minimum in $R_s(H_0)$ to either low and high fields and can reduce $R_s$ below that of an ideal surface. 
The theory shows that the behavior of $R_s(H_0)$ changes as the temperature and the RF frequency are increased, and the field dependence of $Q(H_0)$ can be very sensitive to the materials processing. Our results suggest that the nonlinear RF losses can be minimized by tuning pairbreaking effects at the surface using impurity management or surface nanostructuring. 
\end{abstract}


\maketitle


\section{Introduction}\label{section_introduction}

The physics of electromagnetic response of superconductors, and the fundamental limits dissipation in the Meissner state at low temperatures and frequencies has recently attracted much interest. The issue of ultra-low-dissipation is particularly important for  microresonators for quantum computing or radio-frequency (rf) superconducting cavities for particle accelerators ~\cite{Zmuidzinas, Devoret, 2002_Hein,  2012_Gurevich_review, Padamsee_book}. 
At temperatures $T$ well below the critical temperature $T_c$ and frequencies $\omega$ smaller than the gap frequency $2\Delta/\hbar$, $s$-wave superconductors have very small surface resistance $R_s\propto \exp(-\Delta/T)$ ~\cite{1958_Mattis_Bardeen}. 
Indeed, the Nb cavities typically have $R_s \sim 10~ \rm n \Omega$ at $ 2\,{\rm K}$ and $ 1\,{\rm GHz}$,  
which translates into huge quality factors $Q \propto 1/R_s \sim 10^{10}$-$10^{11}$.  
The surface resistance depends on the amplitude $H_0$ of the rf magnetic field $H(t)=H_0\sin\omega t$ and can be 
significantly altered by the materials treatments. 
For instance, $R_s$ of electropolished Nb cavities ~\cite{Padamsee_book} at $2\,{\rm K}$ and $ 1\,{\rm GHz}$ increases with the rf field amplitude, consistent with the reduction of a quasiparticle gap and the superfluid density by the rf pairbreaking currents ~\cite{parment,bardeen,kopnin}. This manifests itself in a field-dependent London penetration depth $\lambda(H)$ and the nonlinear Meissner effect ~\cite{nme1,nme2,nme3,nme4,nme5,nme6}. 
Yet the Nb cavities infused with nitrogen ~\cite{2013_Grassellino}, titanium ~\cite{2013_Dhakal} or other impurities ~\cite{cornell,jlab} can exhibit a striking field-induced {\it reduction} of $R_s(H_0)$ by factors of 2-4 as $H_0$ increases from $0$ to $ \lesssim 0.5 H_c$, where $H_c$ is the thermodynamic critical field. 

The behavior of $R_s(H_0)$ is determined by multiple mechanisms including interplay of temporal oscillations of the density of states (DOS)  and the kinetics of nonequilibrium quasiparticles under the strong rf field. The field dependence of $R_s(T,H_0)$ is also sensitive to the ratios $\omega/T$ and $T/\Delta$, as well as the electronic structure and compositional inhomogeneities at the surface. It was shown~\cite{2014_Gurevich_PRL} that the well-known effect of broadening of the DOS gap peaks by the pairbreaking current ~\cite{2003_Anthore_PRL, 1964_Maki, 1965_Fulde, 1969_Maki}  can result in a pronounced minimum in $R_s(H_0)$, in agreement with experiment ~\cite{2013_Dhakal,2014_Ciovati_APL}. Such microwave reduction of the surface resistance ~\cite{2014_Gurevich_PRL,2017_Gurevich_SUST} is a manifestation of a general effect by which $R_s$ can be reduced by engineering an optimum broadening of the DOS peaks at the surface using  {\it pairbreaking} mechanisms. These mechanisms can be due to the rf Meissner currents, magnetic impurities, local reduction of the pairing constant, or a proximity-coupled normal layer which models nonstoichiomentry and metallic suboxide layers at the surface ~\cite{2017_Gurevich_Kubo}. 

Sparse magnetic impurities in Nb are among realistic materials features ~\cite{2005_Casalbuoni, 2011_Proslier} which broaden the DOS peaks~\cite{1966_Fulde_Maki, 1969_Maki, 2006_Balatskii, 2011_Fominov, 2012_Kharitonov} and can reduce the low-field $R_s$ by $\sim 50\%$ despite a small reduction of $T_c$ ~\cite{2017_Gurevich_Kubo}.  The DOS broadening and the appearance of subgap states at quasiparticle energies $|\epsilon| < \Delta$ have been revealed by numerous tunneling experiments ~\cite{2003_Zasadzinski, 2013_Dhakal}.  Such DOS has been commonly described by the phenomenological Dynes model which incorporates a constant quasiparticle lifetime $\hbar/\Gamma$ by the replacement $\epsilon \to \epsilon+i\Gamma$ ~\cite{1978_Dynes, 1984_Dynes}:
\begin{equation}
N(\epsilon)=\mbox{Re}\frac{N_s(\epsilon+i\Gamma)}{\sqrt{(\epsilon+i\Gamma)^2-\Delta^2}},\quad \epsilon > 0,
\label{Nd}
\end{equation} 
where $N_s$ is the density of states in the normal state.  Numerous STM experiments have shown that the DOS broadening can be significant, particularly in thin films and bilayers \cite{blstm1,blstm2,stmPb}. The subgap states have been attributed to inelastic scattering of quasiparticles on phonons \cite{inelast,kopnin}, Coulomb correlations \cite{coulomb}, anisotropy of the Fermi surface \cite{anis}, inhomogeneities of the BCS pairing constant \cite{larkin}, magnetic impurities \cite{2006_Balatskii}, spatial correlations in impurity scattering \cite{2006_Balatskii,meyer}, or diffusive surface scattering. \cite{arnold} 

The Dynes model with a constant $\Gamma$ has not been derived from a microscopic theory (see, e.g., Ref. \onlinecite{feigel} for an overview of different mechanisms), yet Eq. (\ref{Nd}) gives an insight into how the DOS broadening could affect $R_s(T)$ at different temperatures.  For instance, at $(\omega,\Gamma)\ll T\ll \Delta$, the surface resistance is mostly determined by thermally-activated quasiparticles with $\epsilon\approx\Delta$. In this case the DOS broadening {\it reduces} $R_s(T)\propto e^{-\Delta/T}\ln (T/\Gamma)$ ~\cite{2017_Gurevich_SUST}.  However, the effect of subgap states reverses at very low temperatures $T< (\Gamma,\omega)$ for which $R_s(T)$ is dominated by quasiparticles with $\epsilon\ll\Delta$ if $\Gamma(\epsilon)$ does not vanish at $\epsilon\to 0$. In this case a finite DOS at the Fermi level $N(0)=N_s\Gamma/\Delta$ increases $R_s$ as compared to the BCS model, giving rise to a residual surface resistance $R_i \propto (\Gamma/\Delta)^2$ at $T\to 0$ ~\cite{2017_Gurevich_SUST, 2017_Gurevich_Kubo}. This dependence of $R_s$ on $\Gamma$ can be used to minimize $R_s(T)$ in a particular temperature region by engineering an optimum DOS at the surface.  While ways of affecting $\Gamma$ in the bulk are not well understood, a better way of engineering the optimum DOS is to use the materials treatment and nanostructuring of the surface of superconducting materials which usually have a thin layer of weakened superconductivity.  
For instance, the Nb surface is covered by a layer of dielectric ${\rm Nb_2 O_5}$  oxide followed by a few nm thick layer of normal metallic suboxides. 
Other materials such as ${\rm Nb_3 Sn}$, ${\rm MgB_2}$, or iron-based superconductors can exhibit a significant surface nonstoichiometry, which can be modeled by a thin normal (N) layer coupled with the bulk by the proximity effect or a superconducting (S$^\prime$) layer separated by a thin insulating (I) layer from the S substrate \cite{mlag1,mlag2,mltk1,mltk2,mlc}. Modification of realistic surface structures can cause profound changes in the low-energy DOS which determines the surface resistance. In particular, it has been shown ~\cite{1989_Golubov, 1996_54_Belzig, 2017_Gurevich_Kubo} that a proximity-coupled N layer causes a disturbance of DOS at $\epsilon\approx \Delta$ which extends into the S region over distances much greater than the coherence length.  As a result, a thin N layer with a moderately transparent NS interface can reduce $R_s$ at weak RF fields  by $\sim 15\%$ relative to the ideal surface~\cite{2017_Gurevich_Kubo}. Here both the thickness of N layer or the transparency of the N-S interface can be tuned by the materials processing and heat treatment. 

Nonlinear dc screening for an ideal surface ~\cite{nme1,nme2,nme3,nme4,nme5,nme6} and proximity-coupled N-S sandwiches, including the superconductivity breakdown under a strong dc magnetic field ~ \cite{1996_53_Belzig,1997_Fauchere,1989_Belzig,2003_Galaktionov} have been thoroughly investigated in the literature. Yet the surface resistance, which is rather sensitive to the details of DOS at $|\epsilon |<\Delta$, paribreaking effects and nonequilibrium kinetics of quasiparticles, has been understood to a much lesser extent.  Recently we calculated $R_s$ for a proximity-coupled N layer under a weak rf field ~\cite{2017_Gurevich_Kubo} but a theory of the field-dependent nonlinear surface resistance in the Meissner state affected by different pairbreaking effects at the surface is lacking. Such a theory should take into account the microwave reduction of $R_s$ due to current pairbreaking ~\cite{2014_Gurevich_PRL,2017_Gurevich_SUST}, the nonlinear rf response of a proximity-coupled N layer and nonequilibrium kinetics of quasiparticles due to collision with phonons and impurities under strong rf field \cite{kopnin,kramer}. In this work we address some of these issues and calculate the field-dependent surface resistance for an imperfect surface using the Usadel equations for dirty superconductors ~\cite{1970_Usadel, 1999_Belzig_review, 2004_Golubov_review}. Our results show how the dependence of $R_s(H_0)$ on the rf amplitude is affected by multiple realistic materials features and suggest ways by which $R_s(H_0)$ could be optimized by tuning the concentration of paramagnetic impurities or properties of the N layer. These results may be useful for improving the quality factors of the resonant cavities for particle accelerators and microresonators for quantum information processing and photon detectors.

The paper is organized as follows. In Sec. II, the geometry of the problem, the Usadel equations, and 
the boundary conditions are formulated. We obtain formulas for a low-frequency nonlinear surface resistance and evaluate a contribution  
of nonequilibrium effects.
In Sec. III, we calculate $R_s(H_0)$ for a superconductor with paramagnetic impurities and show that they can produce a significant minimum in $R_s(H_0)$ as a function of $H_0$.  In Sec. IV, we investigate the effect of the Dynes parameter $\Gamma$ on $R_s$ and show that a finite $\Gamma$ results in a residual resistance at $T\ll T_c$ and  affects a nonlinear field dependence of $R_s(H_0)$ in the way similar to that of paramagnetic impurities at intermediate temperatures. 
In Sec. V, we consider the effect of SIS$^\prime$ surface nanostructuring which can shift the minimum in $R_s(H_0)$ to higher fields.
In Sec. VI, we calculate $R_s(T,H_0)$ for a superconductor covered with a thin, proximity-coupled N layer and investigate the field dependence of $R_s(H_0)$ and the position of the minimum of $R_s(H_0)$ as functions of temperature, thickness of the N layer and the contact resistance $R_B$. 
In Sec. VII, we discuss implications of our results for engineering an optimum DOS at the surface to reduce the rf losses in superconductors under strong electromagnetic fields.

\section{Green's functions and surface resistance}

\subsection{Usadel equation}

\begin{figure}[tb]
   \begin{center}
   \includegraphics[width=0.95\linewidth]{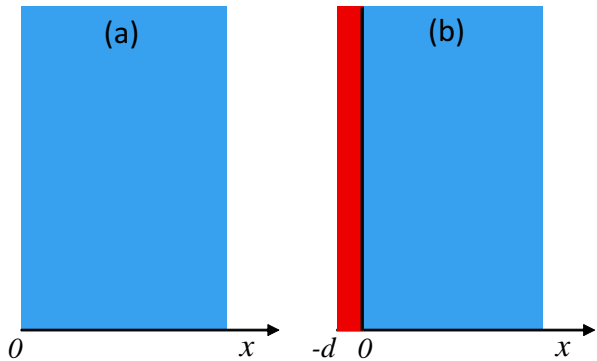}
   \end{center}\vspace{0cm}
   \caption{
(a) A superconductor with an ideal surface. 
(b) A superconductor covered with a proximity-coupled N layer of thickness $d$, or an S$^{\prime}$-I-S multilayer structure. 
The vertical black line shows either the S-N interface with the contact resistance $R_B$ or a thin dielectric (I) layer. 
   }\label{fig1}
\end{figure}

Consider the geometry shown in Fig.~\ref{fig1} which represents a superconductor with an ideal surface or a superconductor S covered with a proximity-coupled $N$ layer or a layer of another superconductor $S^\prime$  at $-d \le x <0$.  For a dirty superconductor, the equilibrium normal and anomalous Green's functions $G=\cos\theta$ and $F=\sin\theta$ satisfy the thermodynamic Usadel equation:
\begin{eqnarray}
\frac{D}{2} \theta'' = s \sin\theta \cos\theta + \tilde{\omega}_{n} \sin\theta - \Delta(x) \cos\theta , \label{eq:usadel} 
\end{eqnarray}
Here $D$ is the electron diffusivity, 
the prime denotes differentiation with respect to $x$, $\omega_n = \pi T (2n+1)$ are the Matsubara frequencies,
$\tilde{\omega}_{n} = \omega_n +  \Gamma$, 
and $\Gamma$ accounts for a finite quasiparticle lifetime in the Dynes model. 
The pairbreaking parameter $s = DQ^2/2 + \Gamma_p$ contains two contribution: the first term comes from the effect of  
pairbreaking currents and the second term accounts for spin-flip scattering on magnetic impurities. 
Here ${\bf Q}=\nabla\theta +2\pi{\bf A}/\phi_0$ is the gauge invariant phase gradient, $\textbf{A}$ is the vector potential,
$\phi_0$ is the flux quantum, 
and $\Gamma_p$ is the spin-flip parameter due to magnetic impurities \cite{2006_Balatskii, 2011_Fominov, 2012_Kharitonov}. 
For the planar geometry shown in Fig.~\ref{fig1}, 
$(\phi_0/2\pi)Q(x,t) =  -\mu_0 \lambda H_0 e^{-x/\lambda} \sin \omega t $ at $x\ge 0$ and $Q(x,t) = Q(0,t)$ at $x<0$, where 
$\lambda$ is the bulk London penetration depth.  We assume that $H_0$ is well below the superheating field \cite{hs1,hs2,hs3} so  
a weak dependence of $\lambda$ on $H_0$ due to the nonlinear Meissner effect \cite{nme1,nme2,nme3,nme4,nme5,nme6} is negligible. We also neglect    
the field attenuation in the N surface layer with the thickness $d \ll \xi \ll \lambda$. Hence,
\begin{gather}
s = s_0e^{-2x/\lambda} \sin^2 \omega t + \Gamma_p , \label{eq:s} \\
s_0 = \frac{\Delta}{\pi} \biggl( \frac{H_0}{H_c} \biggr)^2, \label{eq:zeta}
\end{gather}
where $H_c = (N_s/\mu_0)^{1/2}\Delta$ is the thermodynamic critical field, 
$\Delta$ is the pair potential at $T=0$, and $\Gamma_p$ is assumed uniform. 
Here $\Delta(x,T,s)$ satisfies the BCS self-consistency equation, 
\begin{eqnarray}
\Delta(x) = 2\pi T g \sum_{\omega_n>0}^{\Omega} \sin\theta(x).
\label{eq:self-consistency}
\end{eqnarray}
Here $g$ is the pairing constant in the S region, 
and the summation over $\omega_n$ is cut off at the Debye frequency $\Omega$. 

Eqs.~(\ref{eq:usadel})-(\ref{eq:self-consistency}) are supplemented by the following boundary conditions at $x=-d$ and $x=\infty$: 
\begin{equation}
\theta'(-d)=0 ,   \qquad
\theta(\infty)= \theta_s , \label{eq:BCsurface}
\end{equation}
Here $\theta_{s}$ defines the bulk Green's function satisfying Eq.~(\ref{eq:usadel}) with $\theta'' \to 0$. 
We also use the standard boundary conditions at the N-S interface:~\cite{1988_Kuprianov_Lukichev}
\begin{eqnarray} 
&&\sigma_n R_B \theta_{-}' = \sin (\theta_{0} - \theta_{-}) , 
\label{eq:BC1} \\
&&\sigma_n \theta_{-}' = \sigma_s \theta_{0}' ,
\label{eq:BC2} 
\end{eqnarray}
Here $\theta_{-} \equiv \theta|_{x=-0}$, 
$\theta_{0} \equiv \theta|_{x=+0}$, 
$R_B$ is the N-S contact resistance, 
and $\sigma_n$ and $\sigma_s$ are the normal-state conductivities in the N and S regions, respectively. 
It is convenient to define the dimensionless parameters: 
\begin{eqnarray}
&&\alpha = \frac{N_{n}}{N_{s}} \frac{d}{\xi_{s}} , \label{eq:alpha} \\
&&\beta = \frac{4e^2}{\hbar} R_B N_{n} \Delta d , \label{eq:beta}
\end{eqnarray}
where $N_{n}$ and $N_{s}$ are the normal densities of states in N and S regions,   
and $\xi_{n}=\sqrt{D_{n}/2\Delta}$ and $\xi_{s}=\sqrt{D_{s}/2\Delta}$ are the respective coherence lengths. 
More general boundary conditions for quasiclassical Green's functions are given in Refs.~\onlinecite{1999_Nazarov, 2015_Eschrig}. Yet  
Eqs.~(\ref{eq:BC1}) and (\ref{eq:BC2}) can be used if the NS interface has a small transmission coefficient $t \sim d/\beta \xi_0  \ll 1$ namely $d/\xi_0 \ll \beta < \infty$, 
where $\xi_0$ is a coherence length in the clean limit~\cite{2017_Gurevich_Kubo}. 
For a thin N layer with $d \ll \xi_{s} < \xi_0$, 
the condition $t\ll 1$ includes the essential cases with both $\beta >1$ and $\beta \ll 1$. 

Calculation of the electromagnetic response requires retarded Green's functions $G^R = \cosh\theta$ and $F^R = \sinh\theta$. In the case of dc currents and magnetic fields, $\theta$ satisfies the real-frequency Usadel equation:  
\begin{eqnarray}
\frac{D}{2} \theta'' = s\sinh\theta \cosh\theta - i \tilde{\epsilon} \sinh\theta + i\Delta(x) \cosh\theta ,  \label{eq:R_Usadel}
\end{eqnarray}
where $\tilde{\epsilon} = \epsilon + i\Gamma$. The quasiparticle DOS is given by 
\begin{eqnarray}
n(\epsilon) = \frac{N(\epsilon)}{N_s} = {\rm Re}\, G^R (\epsilon).
\label{doss}
\end{eqnarray}
For a uniform superconductor with $s=0$, $G^R=\tilde{\epsilon}/\sqrt{\tilde{\epsilon}^2-\Delta^2}$ and Eq. (\ref{doss}) 
reduces to Eq. (\ref{Nd}).    

The pair potential $\Delta_s$ for an ideal surface in the case of weak pairbreaking ($s\ll \Delta, \Gamma\ll\Delta$) was calculated in Appendix \ref{appendix:Delta}:
\begin{equation}
\Delta_s=\Delta-\Gamma-\frac{\pi s}{4}, \qquad T\ll T_c,
\label{Deltsg}
\end{equation}
where $\Delta=2\Omega\exp(-1/g)$.

\subsection{Surface resistance} \label{section:Rs}
We consider here type-II superconductors $(\lambda\gg\xi)$ for which the nonlocal BCS electromagnetic response~\cite{1958_Mattis_Bardeen} simplifies to the local relation $\textbf{J}_\omega=\sigma(\omega)\textbf{E}_\omega$ between the Fourier components 
of the current density $\textbf{J}_\omega$ and the electric field $\textbf{E}_\omega$. Here $\sigma(\omega)=\sigma_1-i\sigma_2$ is a complex conductivity, where
$\sigma_2(\omega)=1/\mu_0\omega\lambda^2$ accounts for the Meissner effect, and the quasiparticle conductivity
$\sigma_1(\omega)$ determines the rf dissipation. To express $R_s$ in terms of $\sigma_1$,  we use $E_\omega(x) =-i\omega A_\omega(x)$ and $J_\omega(x) = -i\omega \sigma(x,\omega) A_\omega(x)$ and calculate the power dissipation per unit surface $R_s H_0^2/2 = (1/2) \int_0^\infty {\rm Re} [E J^*] dx$ by  integrating the local power density $\sigma_1E^2(x)$. For the case shown in Fig.~\ref{fig1}(a,b), this yields
\begin{eqnarray}
R_s &=& \omega^2 \mu_0^2 \lambda^2 
\biggl[ 
\int_{-d}^{0} \!\!dx \sigma_1(x) +
\int_{0}^{\infty} \!\!dx \sigma_1(x) e^{-\frac{2x}{\lambda}} \biggr],
 \label{eq:Rs}
\end{eqnarray}
where $d\ll\lambda$. We used Eq. (\ref{eq:Rs}) previosly to calculate a low-field $R_s$ in a superconductor with a 
proximity-coupled N layer at the surface ~\cite{2017_Gurevich_Kubo}. 

Extension of Eq. (\ref{eq:Rs}) to high rf fields requires taking into account nonlinearities of the electromagnetic response in both $\sigma_1$ and $\sigma_2=(\mu_0\omega\lambda^2)^{-1}$. The dependence of $\lambda$ on $H_0$ usually referred to as the nonlinear Meissner effect \cite{nme1,nme2,nme3,nme4,nme5,nme5}, is rather weak at  $T\ll T_c$, $\omega\ll\Delta$, and $H_0\lesssim 0.5H_c$, so it will be neglected in this work. By contrast $\sigma_1$ controlled by the current-induced DOS broadening and nonequilibrium kinetics of quasiparticles is far more sensitive to a low-frequency rf field than $\sigma_2$ determined by the net superfluid density.  Calculation of the field-dependent $\sigma_1$ in a dirty limit can be done using the time-dependent Usadel equation and kinetic equations for quasiparticles under strong low-frequency rf field \cite{kopnin,kramer}. In this case Eq. (\ref{eq:Rs}) can describe the field-dependent surface resistance, if the quasiparticle conductivity $\sigma_1(H_0)$ is averaged over the rf period, taking into account temporal oscillations of the DOS and the distribution function of quasiparticles caused by the current pair-breaking parameter $s(x,t)$. The nonlinear conductivity $\sigma_1(H_0)$ derived in Appendix B  is given by \cite{2014_Gurevich_PRL}:
\begin{equation}
\sigma_1 = \frac{\sigma_n}{\pi} 
\int_0^{\pi/\omega} \!\!\!\!\! dt 
 \int_{-\infty}^{\infty} \!\!\!\!\! d\epsilon \, [ f(\epsilon,t ) - f(\epsilon + \omega,t) ] M(\epsilon, \omega, x,t),
 \label{sig1}
\end{equation}
where $f(\epsilon,t)$ is a distribution function of quasiparticles and $M[\epsilon, \omega, x, s(x,t)]$ is a spectral function:
\begin{eqnarray}
\!\!\!\!\!\!\!\!M = {\rm Re} G^R(\epsilon) {\rm Re} G^R(\epsilon+\omega) 
+ {\rm Re}F^R (\epsilon) {\rm Re}F^R(\epsilon+\omega). 
\label{M}
\end{eqnarray}
For weak or low-frequency rf fields, $f(\epsilon)$ tends to the equilibrium Fermi distribution $f_0(\epsilon)=(e^{\epsilon/T}+1)^{-1}$, and  
Eq. (\ref{sig1}) takes the form:
\begin{eqnarray}
\!\!\!\!\!\sigma_1 =  \frac{\sigma_n}{\pi} \int_0^{\pi/\omega}\!\!\!\!dt\!\int_{-\infty}^{\infty} \!\! \frac{(1-e^{-\omega/ T})M[\epsilon, \omega, s(x,t)]d\epsilon}{(1+e^{-\epsilon/T}) (e^{\epsilon/T} + e^{-\omega/T}) } . \label{sign}
\end{eqnarray}
This formula determines the local nonlinear conductivity in a type-II superconductor where both the magnitude of the rf field and $\sigma_1(x,H_0)$ vary slowly over $\xi$. 
At low fields $H_0\ll (\omega/\Delta)^{3/4}H_c $ and frequencies $\omega\ll T$, the surface resistance $R_s=\mu_0^2\omega^2\lambda^3\sigma_1/2$ can be calculated from Eqs. (\ref{M}) and (\ref{sign}) using $G^R=\epsilon/\sqrt{\epsilon^2-\Delta^2}$, $F^R=\Delta/\sqrt{\epsilon^2-\Delta^2}$ at $\Gamma=0$ and $d=0$. This yields \cite{Zmuidzinas, 2017_Gurevich_SUST, 1958_Mattis_Bardeen}:
\begin{equation}
R_{\rm MB}  
= \frac{\mu_0^2\omega^2\lambda^3\Delta}{\rho_nT} \ln \biggl( \frac{C T}{\omega} \biggr) e^{-\Delta/T},
\label{eq:Rs_MB}
\end{equation}
where $\rho_n=1/\sigma_n$, and $C\approx 9/2$. 
The logarithmic factor in Eq. (\ref{eq:Rs_MB}) results from two close square root singularities at $\epsilon = \pm\Delta$ and $\epsilon=\pm (\Delta+\omega)$ in $M(\epsilon,\omega)$. 
These singularities characteristic of the idealized BCS model disappear as the DOS peaks are broaden by pair-breaking current~\cite{2014_Gurevich_PRL}, bulk subgap states \cite{2017_Gurevich_SUST} or realistic surface features~\cite{2017_Gurevich_Kubo}. In turn, strong rf fields can drive quasiparticles out of equilibrium, so $f(\epsilon)$ in Eq. (\ref{sig1}) should in general be calculated by solving a kinetic equation \cite{kopnin,kramer}.  

\subsection{Nonequilibrium effects}

\begin{figure}[tb]
   \begin{center}
   \includegraphics[width=1\linewidth]{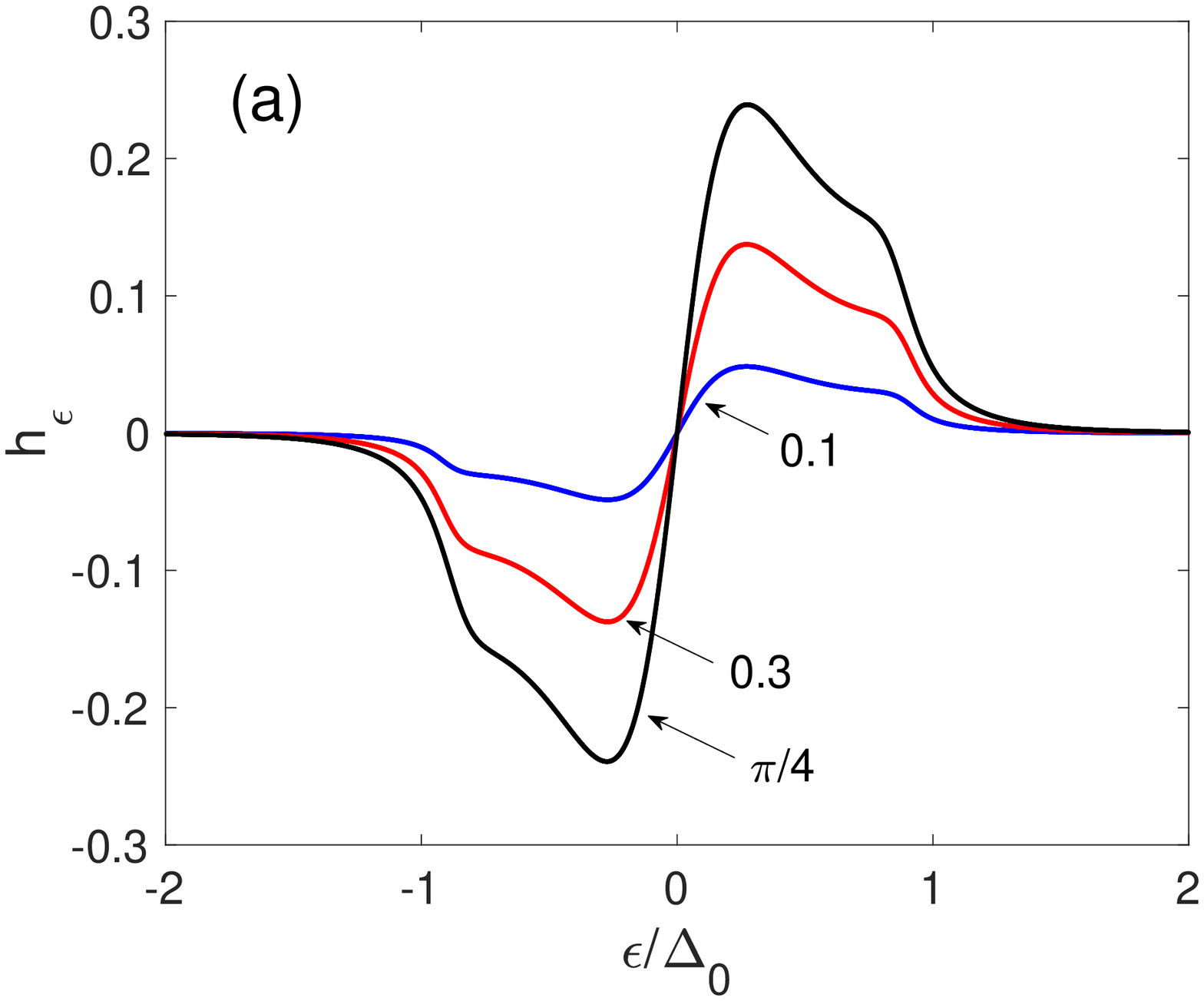}
   \includegraphics[width=1\linewidth]{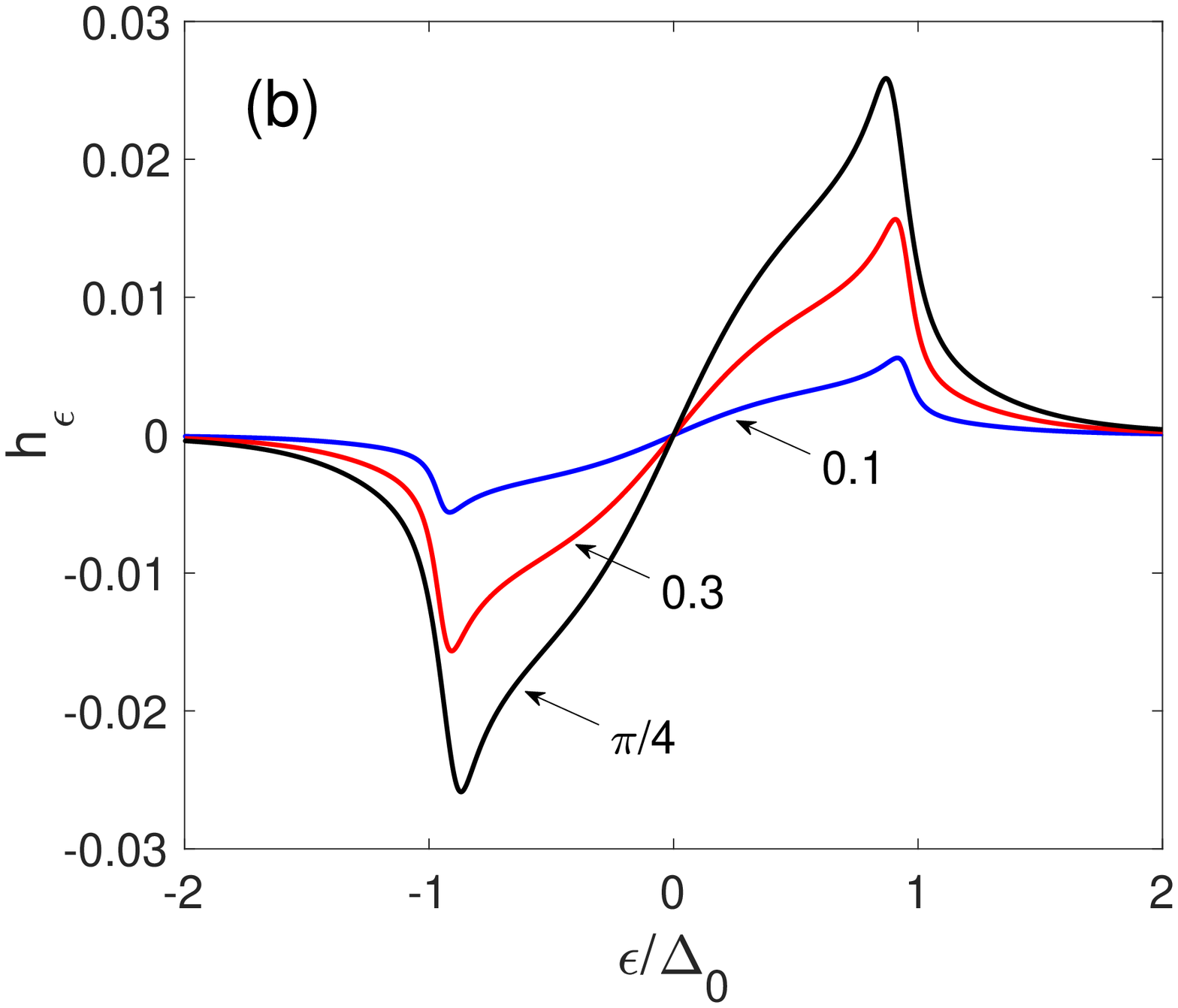}
   \end{center}\vspace{-2mm}
   \caption{
The nonequilibrium amplitude $h_\epsilon$ calculated from Eq. (\ref{h1}) at low frequencies $\omega\ll\gamma_\epsilon$ for:
(a) Low temperatures and moderate DOS broadening.
(b) Intermediate temperatures and weak DOS broadening,
The curves represent Nb at: (a) 2 K, $\Gamma=0.1\Delta$ and (b) 4 K, $\Gamma=0.05\Delta$. 
Here $\Delta=18$ K and $b=0.004$ meV$^{-2}$ from Ref. \onlinecite{kaplan}, $H_0=0.25H_c$, and $\omega=\Delta/400$. The numbers at the curves correspond to the instant values of $\omega t$.
   }\label{fig2}
\end{figure}

\begin{figure}[tb]
   \begin{center}
   \includegraphics[width=1\linewidth]{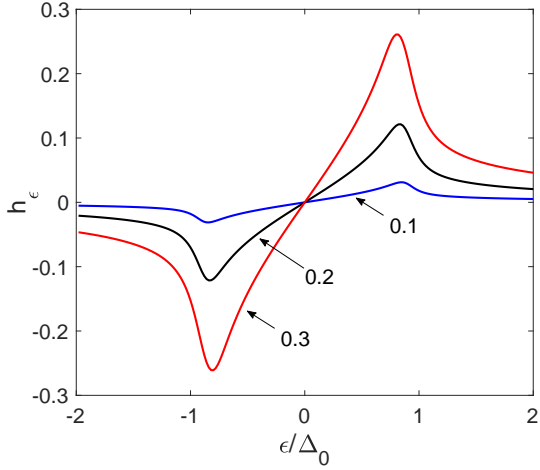}
   \end{center}\vspace{-1mm}
   \caption{
The nonequilibrium amplitude $h_\epsilon$ calculated from Eq. (\ref{h2}) in the high-frequency limit $\omega\gg\gamma_\epsilon$ for 
Nb at 2 K, $\Gamma=0.1\Delta_0$, and different ratios of $H_0/H_c$.
   }\label{fig3}
\end{figure}

Deviation of $f(\epsilon,x,t)$ from $f_0(\epsilon)$ is determined by an absorbed rf power and a rate of the power transfer from quasiparticles to phonons. At $T\ll\Delta$ the power transfer bottleneck is provided by scattering of quasiparticles on phonons which determine an inelastic scattering time $\tau_s$ and a  recombination time of Cooper pairs $\tau_r$, where $\tau_s$ and $\tau_r$ depend on $T$ and $\epsilon$~ \cite{kopnin,kaplan}. For instance, $\tau_r$ and $\tau_s$ at $\epsilon=\Delta$ in the absence of the rf field are given by~\cite{kaplan}
\begin{eqnarray}
\tau_r = \tau_1 \biggl( \frac{T_c}{T} \biggr)^{1/2} e^{\Delta/T}, 
\hspace{1cm}
\tau_s = \tau_2 \biggl( \frac{T_c}{T} \biggr)^{7/2}, 
\label{tauu}
\end{eqnarray}
where $\tau_1$ and $\tau_2$ are materials constants. For Nb with $T_c=9.2\,{\rm K}$, $\Delta=1.9T_c$, $\tau_1 = 3\cdot 10^{-12}\,{\rm s}$, and $\tau_2 = 8\cdot 10^{-11}\,{\rm s}$ ~\cite{kaplan}, Eq. (\ref{tauu}) gives $\tau_r \sim 4\cdot 10^{-8}\, {\rm s}$ and $\tau_s \simeq 2\cdot 10^{-8}\,{\rm s}$ at $T=2\,{\rm K}$. As $T$ increases to 4K, the time constants $\tau_r\simeq 4\cdot 10^{-10}\,{\rm s}$ and $\tau_s\simeq2\cdot 10^{-9}\,{\rm s}$ diminish but remain much longer than the relaxation time constant of the superconducting condensate, $\hbar/\Delta\sim 10^{-12}$~s. At $T\ll T_c$ and $\omega\ll \Delta/\hbar$ an exponentially small density of quasiparticles has practically no effect on the dynamics of the condensate which reacts nearly instantaneously to the time-dependent currents. As a result, the spectral function $M$ in Eq. (\ref{M}) is determined by the quasi-static Green's functions. However, $\sigma_1$ can be very sensitive to a slow dynamics of nonequilibrium quasiparticles which control the distribution function $f$. Depending on the magnitude of the rf field, and the relation between $\omega$ and the relaxation time constants, quasiparticles can either follow the temporal variations of $M(t)$ if $(\tau_r,\tau_s)\omega \gg 1$ or relax quickly to the equilibrium state if $ (\tau_r,\tau_s)\omega\ll 1$.  Both $\tau_r$ and $\tau_s$ increase strongly as $T$ decreases, so nonequilibrium effects become more pronounced at $T\ll T_c$. 

We evaluate non-equilibrium effects in $f$ using the kinetic equation for the matrix distribution function in a dirty superconductor in a low-frequency 
magnetic field \cite{kopnin,kramer,lo}. In this case the quasiparticle electron-hole symmetry and the charge neutrality are preserved and the matrix kinetic equation reduces to only one equation for a correction to the odd in $\epsilon$ distribution function $\delta f_\epsilon(t)$. It is convenient to define
\begin{equation}
\delta f_\epsilon(x,t)=\frac{h_\epsilon(x,t)}{\cosh^2(\epsilon/2T)},
\label{he}
\end{equation}
where the amplitude $h_\epsilon(x,t)$ quantifies deviation of $f$ from equilibrium (nonequilibrium effects are negligible if $h_\epsilon\ll 1$). 
Using Eq. (\ref{Deltsg}) and the parameterization $G^R=\cosh(u+iv)$, $F^R=\sinh(u+iv)$, the kinetic equation obtained 
in Ref. \onlinecite{kramer} can be recast in the form:
\begin{gather}
\cos v_{\epsilon}\cosh u_{\epsilon}\frac{\partial h_{\epsilon}}{\partial t}= D\nabla\cdot(\cos^{2}v_{\epsilon}\nabla h_{\epsilon}) +
\nonumber \\
\frac{1}{2T}\cos v_{\epsilon}\sinh u_{\epsilon}\left(\frac{\pi}{4}-\sin v_{\epsilon}\cosh u_{\epsilon}\right)\frac{\partial s}{\partial t}-I_{ph},
\label{kin1}
\end{gather} 
where $I_{ph}$ is the electron-phonon collision integral:
\begin{gather}
I_{ph}=2\pi b\cosh\frac{\epsilon}{2T}\cos v_{\epsilon}\cdot
\nonumber \\
\!\int_{-\infty}^{\infty}\!d\epsilon'(h_{\epsilon}-h_{\epsilon'})\frac{(\epsilon-\epsilon')^2\cos v_{\epsilon'}\cosh(u_{\epsilon}-u_{\epsilon'})}{\cosh(\epsilon'/2T)\sinh(|\epsilon-\epsilon'|/2T)}.
\label{I}
\end{gather}
Here phonons are assumed to be at equilibrium, and the low-energy phonon spectral function $\mu(\omega)=\alpha^{2}(\omega)F(\omega)= b\omega^{2}$ is used, where the constant $b$ was calculated in Ref. \onlinecite{kaplan} for different materials.  

We neglect the diffusion term in Eq. (\ref{kin1})  assuming that the rf currents vary slowly over $\xi$ and the diffusion length of quasiparticles. We also neglect the contribution of $h_{\epsilon'}$ to the collision integral (\ref{I}) for qualitative evaluation of $h_\epsilon$ (see Ref. \onlinecite{kramer}), so Eq. (\ref{kin1}) simplifies to:
\begin{gather}
\frac{\partial h_\epsilon}{\partial t}+\gamma_\epsilon(t)h_\epsilon=F_\epsilon(t),
\label{kins} \\
\!\!\gamma_\epsilon =2\pi b\frac{\cosh(\epsilon/2T)}{\cosh u_\epsilon}\!
\int_{-\infty}^{\infty}\!\!d\epsilon'\frac{(\epsilon-\epsilon')^2\cos v_{\epsilon'}\cosh(u_{\epsilon}-u_{\epsilon'})}{\cosh(\epsilon'/2T)\sinh(|\epsilon-\epsilon'|/2T)},
\label{gamm} \\
F_\epsilon(t)=\frac{1}{2T}\tanh u_{\epsilon}\left(\frac{\pi}{4}-\sin v_{\epsilon}\cosh u_{\epsilon}\right)\frac{\partial s}{\partial t}.
\label{fas}
\end{gather}
Here $\gamma_\epsilon$ is the rate of electron-phonon inelastic collisions, and $F_\epsilon(t)$ is a driving term. Because $u_\epsilon(t)$ and $v_\epsilon(t)$ depend on $s(t)$, both $\gamma_\epsilon(t)$ and $F_\epsilon(t)$ are functions of time. 

At low frequencies $\omega \ll \gamma_\epsilon$, the term $\partial h/\partial t$ in Eq. (\ref{kins}) can be neglected, and
\begin{equation}
h_\epsilon =\frac{1}{2T\gamma_\epsilon}\tanh u_{\epsilon}\!\left[\frac{\pi}{4}-\sin v_{\epsilon}\cosh u_{\epsilon}\right]\!\frac{\partial s}{\partial t}.
\label{h1}
\end{equation}
At high frequencies $\omega\gg\gamma_\epsilon$, the electron-phonon collisions term in Eq. (\ref{kins}) can be neglected, and
\begin{equation}
h_\epsilon =\frac{1}{2T}\!\int_0^{s(t)}\!\!\!\tanh u_{\epsilon}\left[\frac{\pi}{4}-\sin v_{\epsilon}\cosh u_{\epsilon}\right]\!ds.
\label{h2}
\end{equation}
The functions $h_\epsilon(t)$ given by Eqs. (\ref{h1}) and (\ref{h2}) become of the same order of magnitude at $\omega\sim\gamma_\epsilon$.    

Shown in Figs. \ref{fig2} and \ref{fig3} are the nonequilibrium amplitudes $h_\epsilon$ calculated from Eqs. (\ref{h1}) and (\ref{h2}). Here 
$\theta=u+iv$ was calculated by numerically solving the uniform quasistatic Usadel equation (\ref{eq:R_Usadel}). 
Hence, it follows that: 1. The magnitude of $h_\epsilon$ increases as $T$ decreases, consistent with the above qualitative analysis. 
2. The function $h_\epsilon$ has broadened peaks at $\epsilon \simeq \pm\Delta$ and decreases rapidly at $|\epsilon|>\Delta$ as $|\epsilon|$ increases.  3.  
Nonequilibrium subgap states which appear in $h_\epsilon$ become more pronounced as the rf field increases and $T$ decreases. 4. As the subgap parameter $\Gamma$ increases, the amplitude of the gap peaks in $h_\epsilon$ diminishes. 

As follows from Figs. \ref{fig2} and \ref{fig3}, the nonequilibrium effects become negligible $(h_\epsilon\ll 1)$ as $\omega$ and $H_0$ decrease or $\Gamma$ increases. For instance, at $H_0 \lesssim 0.5 H_c$ and $\omega\simeq (2-3)\cdot 10^{-3}\Delta$ (which corresponds to 1-2 GHz for Nb), the condition $h_\Delta\lesssim 1$ is satisfied at 4K at $\Gamma/\Delta = 0.05$ and at 2K at $\Gamma/\Delta=0.1$. The reduction of $h_\epsilon$ as $\Gamma$ increases reflects acceleration of the energy relaxation of quasiparticles due to a finite DOS at $|\epsilon|<\Delta$ and the smearing out the DOS gap singularity.  In turn, faster energy relaxation and  diminishing nonequilibrium effects are facilitated by a proximity coupled N layer at the surface which provides a source of low-energy quasiparticles with $\epsilon<\Delta$. This layer plays a role of a "quasiparticle trap" which can reduce the density of nonequilibrium quasiparticles in a superconductor \cite{qpt1,qpt2}. Given the wide range of the materials parameters, $H_0$, $\omega$ and $T$ where the nonequilibrium effects are not of prime importance, we assume here the Fermi-Dirac distribution in Eq. (\ref{sig1}) to identify the material features which can be tuned to optimize $R_s(H_0)$.

\section{An ideal surface with magnetic impurities in the bulk}
%
\begin{figure}[tb]
   \begin{center}
   \includegraphics[width=0.95\linewidth]{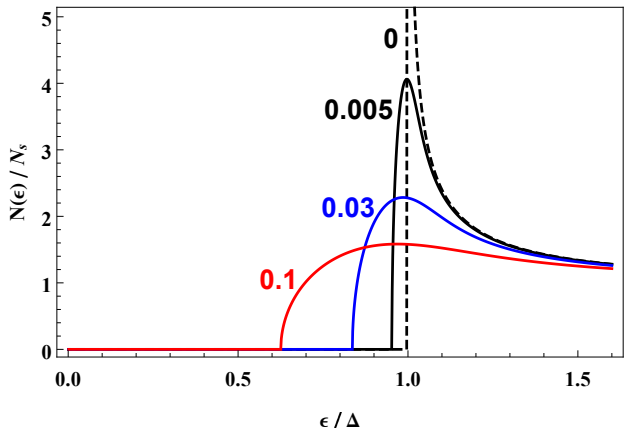}
   \end{center}\vspace{0cm}
   \caption{
Densities of states for $s/\Delta =0, 0.005, 0.03, 0.1$. 
   }\label{fig4}
\end{figure}

Consider the effect of pairbreaking magnetic impurities on the nonlinear surface resistance at $\Gamma=0$. Here $G^R=\cosh\theta$ and $F^R=\sinh\theta$ in the spectral function $M$ are calculated from a quasi-static Usadel equation (\ref{eq:R_Usadel}) for $\theta=u+iv$, where the low-frequency rf currents $(\omega\ll \Delta)$ vary slowly over $\xi$ so $\theta''$ can be neglected:
\begin{eqnarray}
i s \sinh\theta\cosh\theta + \epsilon \sinh\theta - \Delta_s \cosh\theta = 0. 
\label{eq:R_Usadel_Mag}
\end{eqnarray}
Separating real and imaginary parts of Eq. (\ref{eq:R_Usadel_Mag}) gives a quadratic equation for $\sin v$ and a cubic equation for $\sinh 2u$ with the solutions~\cite{2014_Gurevich_PRL}:
\begin{gather}
\sin v = \frac{1}{2s \cosh u} \Bigl( -\Delta_s + \sqrt{\Delta_s^2 - s^2 \sinh^2 2u} \Bigr), \label{eq:R_Usadel_Mag_sol1}  \\
\sinh 2u = \frac{1}{s} \Bigl[ (q +\epsilon s \Delta_s)^{1/3} - (q - \epsilon s \Delta_s)^{1/3} \Bigr]  , \label{eq:R_Usadel_Mag_sol2} \\
q  = \bigl[\epsilon^2 s^2 \Delta_s^2  + \frac{1}{27}(\epsilon^2 - \Delta_s^2 + s^2)^3 \bigr]^{1/2} . \label{eq:R_Usadel_Mag_sol3}
\end{gather}
The density of states $n= {\rm Re} G^R = \cosh u \cos v$ calculated from Eqs. (\ref{eq:R_Usadel_Mag_sol1})-(\ref{eq:R_Usadel_Mag_sol3}) is shown in Fig.~\ref{fig4} for different values of the pairbreaking parameter $s$. Here the DOS  
vanishes at energies $|\epsilon|<\epsilon_g$, where ~\cite{1969_Maki}:
\begin{gather}
\epsilon_g = ( \Delta_s^{2/3} - s^{2/3} )^{\frac{3}{2}}, 
\label{eq:eg} \\
\Delta_s=\Delta_0-\frac{\pi}{4}s.
\label{dels}
\end{gather}
Because of current and paramagnetic pairbreaking effects the quasiparticle gap $\epsilon_g$ is smaller than $\Delta$,  
vanishing at $s>0.43\Delta_s$. If $\Gamma_p < 0.22$ the gap $\epsilon_g$ remains finite even at the superheating field $H_{\rm sh} \simeq 0.84H_c$, consistent with the previous result obtained for $\Gamma_p=0$ ~\cite{nme6}.  
  
\subsubsection{Surface resistance}

%
\begin{figure}[tb]
   \begin{center}
   \includegraphics[width=0.95\linewidth]{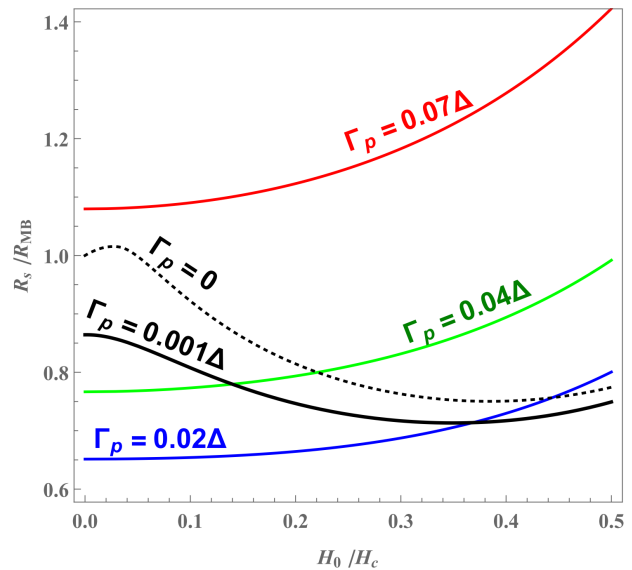}
   \end{center}\vspace{0cm}
   \caption{
$R_s(H_0)$ calculated from Eq.~(\ref{eq:Rs_Mag_2}) at $\omega /\Delta = 0.004$, 
$T /\Delta=0.11$, and $\Gamma_p/\Delta = 0, 0.001, 0.02, 0.04, 0.07$. 
   }\label{fig5} 
\end{figure}
\begin{figure}[tb]
   \begin{center}
   \includegraphics[width=0.95\linewidth]{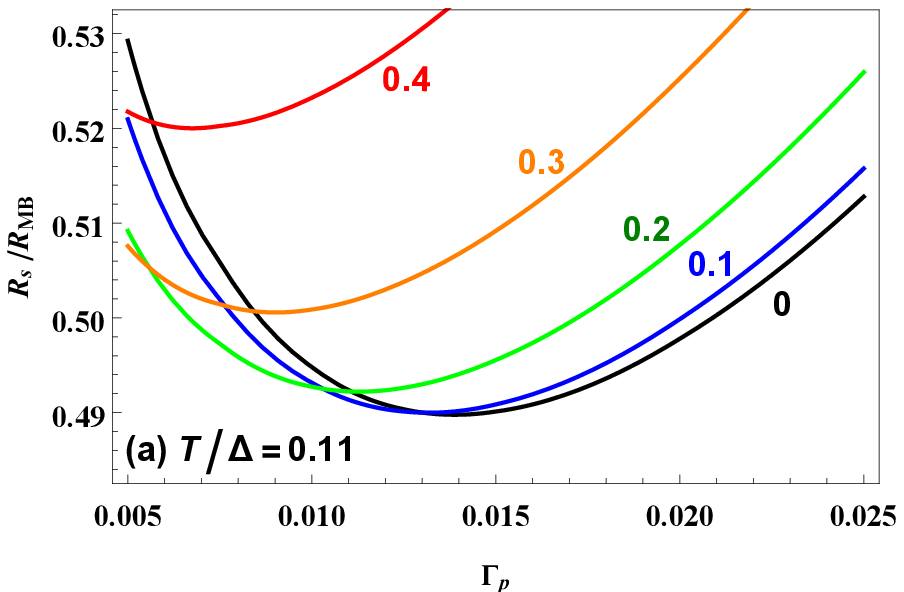}
   \includegraphics[width=0.95\linewidth]{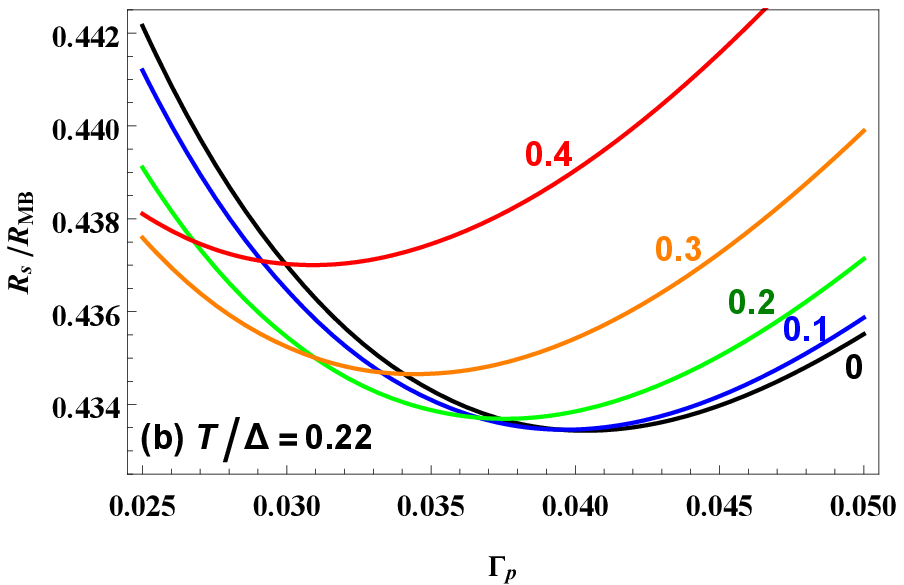}
   \end{center}\vspace{0cm}
   \caption{
$R_s(H_0,\Gamma_p)$ as functions of $\Gamma_p$ calculated for $\omega /\Delta = 0.001$, 
$H_0/H_c = 0, 0.1, 0.2, 0.3, 0.4$ and $T /\Delta =0.11, 0.22$. 
   }\label{fig6}
\end{figure}

We calculate the nonlinear surface resistance $R_s(H_0)$ using Eqs.~(\ref{eq:Rs}), (\ref{M}), (\ref{sign})  
for an ideal surface $(d=0)$. In this case the $x$-integration in Eq. (\ref{eq:Rs}) can be changed to integration over $s$, 
giving
\begin{gather}
R_s = \frac{\mu_0^2\omega^2\lambda^3\Delta}{\pi \rho_n T s_0}\! \int_0^{s_0}\!\! d\zeta \int_0^{\pi}\!\!d\tau\!\int_{\epsilon_g}^{\infty} \!\!\! \frac{M[\epsilon, s(\tau)]d\epsilon}{[1+e^{-\epsilon/T}] [e^{\epsilon/T} + e^{-\omega/T}]},   \label{eq:Rs_Mag_2} \\
s(\tau)=\zeta\sin^2\tau +\Gamma_p,\qquad \zeta =s_0 e^{-2x/\lambda},
\label{sz}
\end{gather}
where $\tau = \omega t$. 
Shown in Fig.~\ref{fig5} is $R_s(H_0)$ calculated from Eqs. (\ref{eq:Rs_Mag_2}) and (\ref{sz}) for different values of $\Gamma_p$, 
where $R_s(H_0)$ is normalized to $R_{\rm MB}$ at $H_0=0$ and $\Gamma_p=0$. 
The dashed curve corresponds to the case of no magnetic impurities $(\Gamma_p=0)$ considered in Ref. \onlinecite{2014_Gurevich_PRL}.

One of the striking features of $R_s(H_0)$ shown in Fig.~\ref{fig5} is a minimum in $R_s(H_0)$ resulting from the interplay of the current-induced broadening of the DOS peaks which reduces $R_s$ and the reduction of $\epsilon_g$ which increases $R_s$ ~\cite{2014_Gurevich_PRL}. This effect of the rf field suppression of $R_s(H_0)$ can be qualitatively understood from Eq. (\ref{eq:Rs_MB}) in which the logarithmic term results from two close square root singularities in $M(\epsilon)$ for the idealized DOS of the BCS model at $\omega\ll T$. This logarithmic singularity at $\omega=0$ is suppressed by either magnetic or current pairbreaking or the Dynes DOS broadening. As was pointed out above, any weak broadening of the DOS peaks reduces $R_s$. For instance, the pairbreaking effects take over if they broaden the DOS peaks to the width  $\delta \epsilon\sim s^{2/3}\Delta^{1/3}$ exceeding $\omega$, so that the gap singularity in $M(\epsilon)$ is cut off at the field-dependent $\delta\epsilon$ rather than $\omega$. Thus, $R_s$ at $s>\omega^{3/2}\Delta^{-1/2}$ can be roughly evaluated by replacing $\omega\to\delta\epsilon$ in Eq. (\ref{eq:Rs_MB}), giving a logarithmic decrease of $R_s$ with $H_0$:
\begin{equation}
R_s\sim \frac{\mu_0^2\omega^2\lambda^3\Delta}{T\rho_n}e^{-\Delta/T}\ln\frac{T \Delta^{-1/3}}{[(H_0/H_c)^{2}\Delta+\Gamma_p]^{2/3}}.
\label{estR}
\end{equation}     
The decrease of $R_s(H_0)$ with $H_0$ eventually stops at the field $H_m$ defined by $\delta\epsilon(H_m)\sim T$. Hence,
\begin{equation}
H_m\sim \left[\frac{T^{3/2}\Delta^{-1/2}-\Gamma_p}{\Delta}\right]^{1/2}H_c.
\label{hm}
\end{equation}
At $H_0=H_m$ and $\omega\lesssim \Gamma_p^{2/3}\Delta^{1/3}$ the surface resistance reaches a minimum which is by the factor $\sim\ln(T/\Delta^{1/3}\Gamma_p^{2/3})$ smaller than $R_s(0)$. As $\Gamma_p$ increases, the minimum in $R_s(H_m)$ gets shallower and shifts to smaller fields. The minimum in $R_s(H_0)$ disappears above a critical concentration of magnetic impurities defining the threshold pairbreaking parameter $\Gamma_p\simeq T^{3/2}\Delta^{-1/2}$ which decreases with $T$. This qualitative consideration is consistent with the numerical calculations of $R_s(H_0)$ done at $T/\Delta=0.11$ and $\omega/\Delta=0.004$. For this case shown in  Fig.~\ref{fig5}, the minimum disappears at $\Gamma_p \gtrsim 0.01\Delta$.

As follows from Fig.~\ref{fig5}, moderate magnetic pairbreaking decreases $R_s(H_0)$ at low fields while increasing 
$R_s(H_0)$ at higher fields. Here $\Gamma_p\simeq 0.01\Delta$ corresponds to an optimum DOS broadening due to magnetic pairbreaking. At higher fields the combined magnetic and current pairbreaking parameter $s$ exceeds the optimal value and $R_s(H_0)$ increases sharply with $H_0$ so the position of the minimum in $R_s(\Gamma_p)$ depends on $H_0$. 
Shown in Fig~\ref{fig6} are $R_s(\Gamma_p)$ curves calculated for various $H_0$ and $T$. 
Here the optimal $\Gamma_p$ proportional to the density of magnetic impurities decreases as $H_0$ increases but increases as $T$ increases. 
For $\Gamma_p = \hbar v_F/2\ell_s \sim 0.01\Delta$, 
the optimum spin-flip scattering mean free path $\ell_s \sim 10^2 \xi_0$ is much larger than $\xi_0$ 
in the absence of low-energy bound states on magnetic impurities~\cite{2017_Gurevich_Kubo}. 

\begin{figure}[tb]
   \begin{center}
   \includegraphics[width=0.95\linewidth]{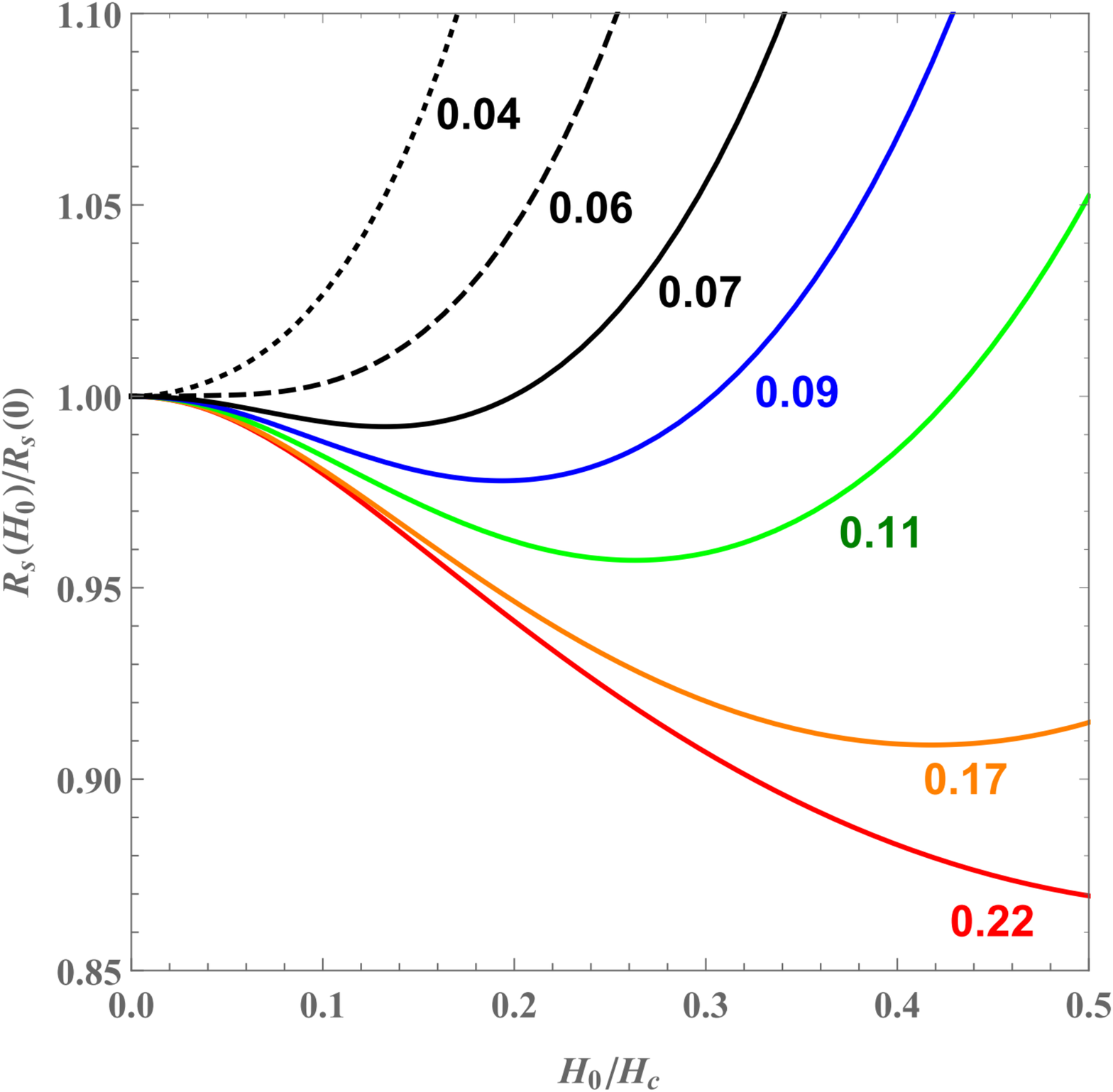}
   \end{center}\vspace{0cm}
   \caption{
$R_s(H_0,T)$ calculated at $\omega/\Delta=0.001$, $\Gamma_p/\Delta=0.005$ and $T/\Delta =0.04, 0.06, 0.07, 0.09, 0.11, 0.17, 0.22$.   
   }\label{fig7}
\end{figure}
\begin{figure}[tb]
   \begin{center}
   \includegraphics[width=0.95\linewidth]{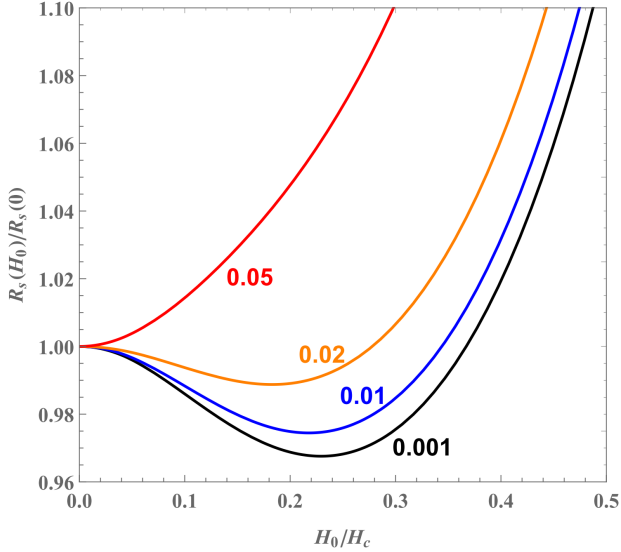}
   \end{center}\vspace{0cm}
   \caption{
$R_s(H_0,\omega)$ calculated at $\Gamma_p=0.005 \Delta$, $T/\Delta= 0.1$ and $\omega /\Delta =0.001, 0.01, 0.02, 0.05$. 
   }\label{fig8}
\end{figure}

The evolution of $R_s(H_0)$ as temperature increases is shown in Fig.~\ref{fig7}. Here 
 the minimum in $R_s(H_0)$ becomes shallower and shifts to lower fields as $T$ decreases. 
This behavior is consistent with the above qualitative discussion around Eqs. (\ref{estR}) and (\ref{hm}) that at low temperatures $T \sim \delta \epsilon \sim \Delta (s/\Delta)^{2/3}$, the effect of DOS broadening on $R_s(H_0)$ weakens and $R_s(H_0)$ is dominated by the reduction of the quasiparticle gap $\epsilon_g(H_0)$ with $H_0$. 
The evolution of $R_s(H_0)$ with the frequency $\omega$ is similar to the effect of $\Gamma_p$ and $T$ considered above. For instance, the results of our calculations presented in Fig.~\ref{fig8} show that the minimum in $R_s(H_0)$ becomes shallower and shifts to lower fields as $\omega$ increases, and disappears at $\omega \gtrsim T$. 
It should be noted that this behavior of $R_s(H_0,T,\omega,\Gamma_p)$ obtained for the equilibrium distribution function of quasiparticle can change at low temperatures where the nonequilibrium effects become important. For instance, the condition of frozen density of quasiparticles caused by slow electron-phonon relaxation $\gamma_\epsilon \ll \omega$ at low $T$ can significantly deepen the minimum in $R_s(H_0)$  as compared to the equilibrium distribution function \cite{2014_Gurevich_PRL,2017_Gurevich_SUST}

\section{An Ideal surface with bulk subgap states}
In this section we address the effect of bulk subgap states on the dependencies of the nonlinear surface resistance on the rf field, frequency and temperature.
\subsection{Green's functions and density of states}

\begin{figure}[tb]
   \begin{center}
   \includegraphics[width=0.95\linewidth]{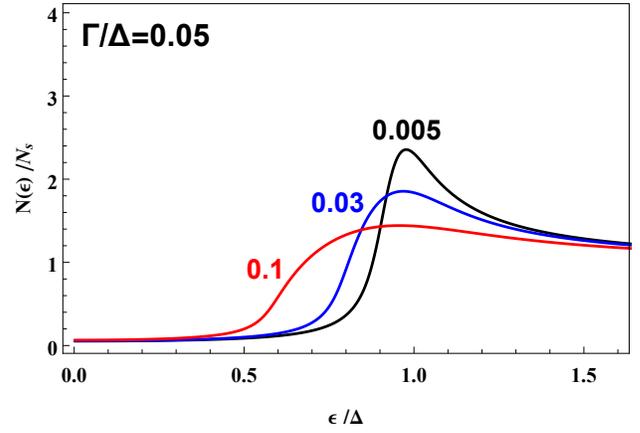}
   \end{center}\vspace{0cm}
   \caption{
Densities of states calculated from Eqs. (\ref{eq:R_Usadel_Gamma_1}) and (\ref{eq:R_Usadel_Gamma_2}) for $\Gamma/\Delta= 0.05$ and $s/\Delta=0.005, 0.03, 0.1$. 
   }\label{fig9}
\end{figure}
Retarded Green's functions are obtained by solving the real-frequency Usadel equation (\ref{eq:R_Usadel_Mag}) with $\epsilon\to \epsilon+ i \Gamma$ and $\Delta_s$ given by Eq. (\ref{Deltsg}).  This uniform Usadel equation for $\theta = u + i v$ yields two coupled equations for $v$ and $u$: 
\begin{gather}
(\epsilon\cosh u - \Delta\sinh u)\sin v+
\nonumber \\ 
\Gamma\sinh u\cos v + \frac{s}{2}\sinh 2u\cos 2v=0,
\label{eq:R_Usadel_Gamma_1} \\
(\Delta\cosh u-\epsilon\sinh u)\cos v+
\nonumber \\
\Gamma\cosh u\sin v+\frac{s}{2}\cosh 2u\sin 2v=0.
\label{eq:R_Usadel_Gamma_2} 
\end{gather}
The density of states $n=\cosh u \cos v$ calculated from Eqs.~(\ref{eq:R_Usadel_Gamma_1}) and (\ref{eq:R_Usadel_Gamma_2}) 
is shown in Fig.~\ref{fig9}. Here the gap peaks in the DOS are smeared even at $s=0$ and subgap states appear at $\epsilon=0$. The density of these subgap 
states increases with $s$. For instance, at $\epsilon=0$, the solution of the uniform Usadel equation in the first order in $(\Gamma,s)/\Delta\ll 1$ gives:
\begin{equation}
n(\epsilon=0) = \frac{\Gamma}{\Delta}\biggl[ 1+ \Bigl( 1+\frac{\pi}{4} \Bigr)\frac{s}{\Delta} \biggr],
\label{noo}
\end{equation}
where Eq. (\ref{Deltsg}) was used. 

\subsection{Surface resistance}

%
\begin{figure}[tb]
   \begin{center}
   \includegraphics[width=0.95\linewidth]{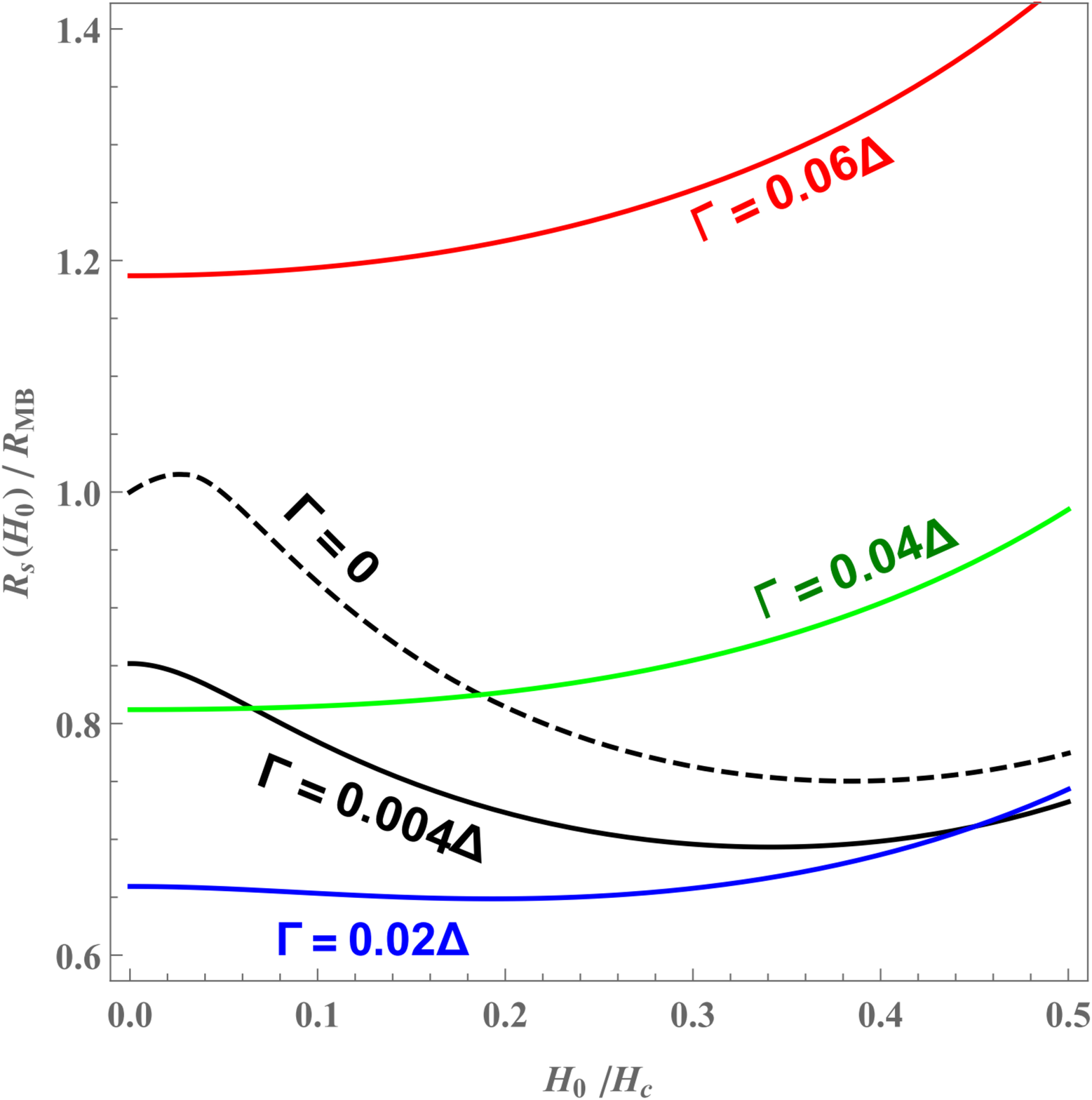}
   \end{center}\vspace{0cm}
   \caption{
$R_s(H_0)$ calculated from Eq.~(\ref{RsG}) at $\Gamma/\Delta = 0, 0.004, 0.02, 0.04, 0.06$, 
$\Gamma_p = 0$, $\omega /\Delta = 0.004$, and $T /\Delta =0.11$.  
   }\label{fig10}
\end{figure}
\begin{figure}[tb]
   \begin{center}
   \includegraphics[width=0.95\linewidth]{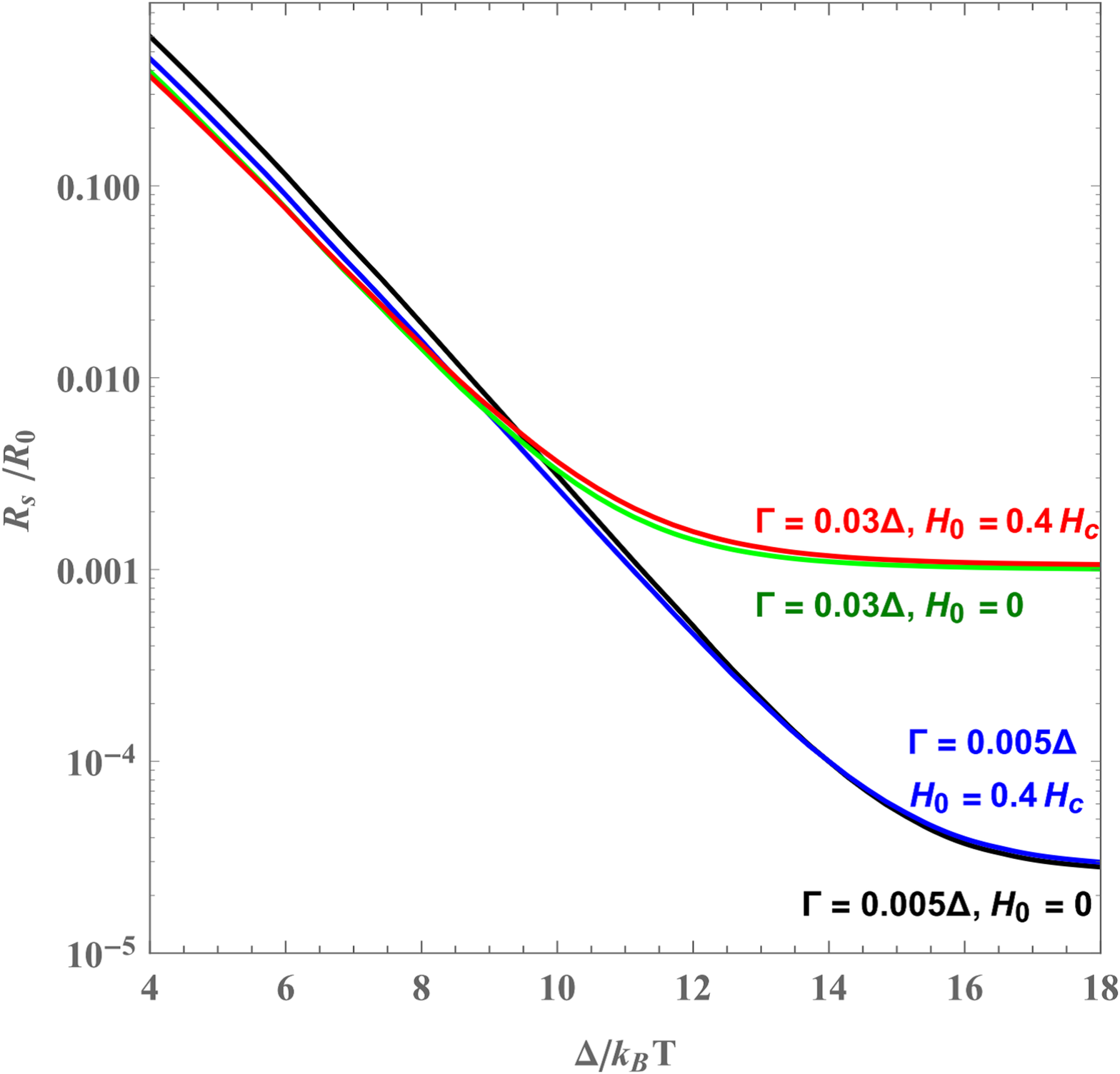}
   \end{center}\vspace{0cm}
   \caption{
Arrhenius plots for $R_s$ calculated at $\omega /\Delta = 0.001$, 
$\Gamma_p = 0$, $\Gamma/\Delta = 0.005, 0.03$, and $H_0/H_c=0, 0.4$. 
Here $R_s$ is normalized by $R_0 = (1/2) \mu_0^2 \omega^2 \lambda^3 \sigma_n$, 
and temperature dependences of $\lambda$ and $\Delta$ at $T<T_c/2$ are neglected. 
   }\label{fig11}
\end{figure}

The surface resistance at $\Gamma>0$ is calculated by generalization of Eq.~(\ref{eq:Rs_Mag_2}): 
\begin{gather}
R_s = \frac{\mu_0^2\omega^2\lambda^3\Delta}{2\rho_n T s_0}\int_0^{s_0}I(\zeta,\omega)d\zeta, 
\label{RsG} \\
I(\zeta,\omega)=\frac{1}{\pi}\int_0^{\pi}\!\!d\tau\!\int_{-\infty}^{\infty} \!\! \frac{M[\epsilon, s(\tau)]d\epsilon}{(1+e^{-\epsilon/T}) (e^{\epsilon/T} + e^{-\omega/T})},
\label{I}
\end{gather}
Here $\tau=\omega t$, and the spectral function $M$ is defined by Eq. (\ref{M}), where $G^R$ and $F^R$ are obtained from the solutions of Eqs. (\ref{eq:R_Usadel_Gamma_1}) and (\ref{eq:R_Usadel_Gamma_2}). 

Shown in in Fig.~\ref{fig10} is $R_s(H_0)$ calculated for different values of $\Gamma$.
Here  the effect of subgap states on the field dependence of $R_s(H_0)$ appears similar to the effect of magnetic impurities shown in Fig.~\ref{fig5}. 
The dashed curve in Fig.~\ref{fig10} corresponds to $R_s(H_0)$ for an ideal surface with no magnetic impurities and bulk subgap states~\cite{2014_Gurevich_PRL}. 
The minimum in $R_s(H_0)$ comes from the interplay of the broadening of the DOS peaks and a reduction of the quasiparticle gap by current. 
Here the DOS peaks are broaden by both rf currents and a finite lifetime of quasiparticle, 
so the minimum in $R_s(H_0)$ shifts to lower fields as $\Gamma$ increases. If 
$\Gamma$ exceeds a critical value $\Gamma_c\sim T^{3/2}\Delta^{-1/2}$ (see Eq. (\ref{hm})), the optimum DOS peak width occurs at $H_0 =0$, and  
$R_s(H_0)$ increases monotonically with $H_0$. 
Yet the effect of subgap states on the temperature dependences of $R_s(H_0,T)$ turns out to be rather different from that of magnetic impurities.

\begin{figure}[tb]
   \begin{center}
   \includegraphics[width=0.95\linewidth]{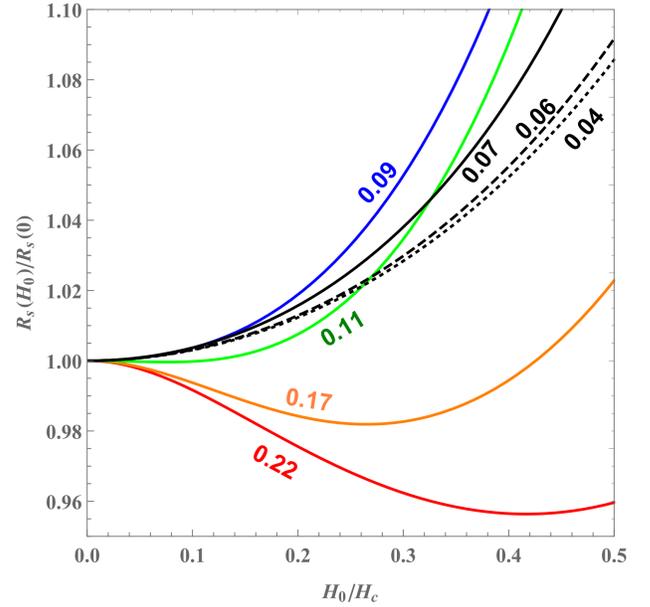}
   \end{center}\vspace{0cm}
   \caption{
$R_s(H_0)$ calculated at $\omega/\Delta = 0.001$, $\Gamma/\Delta=0.03$, 
$T/\Delta= 0.04, 0.06, 0.09, 0.11, 0.17, 0.22$. 
   }\label{fig12}
\end{figure}
\begin{figure}[tb]
   \begin{center}
   \includegraphics[width=0.95\linewidth]{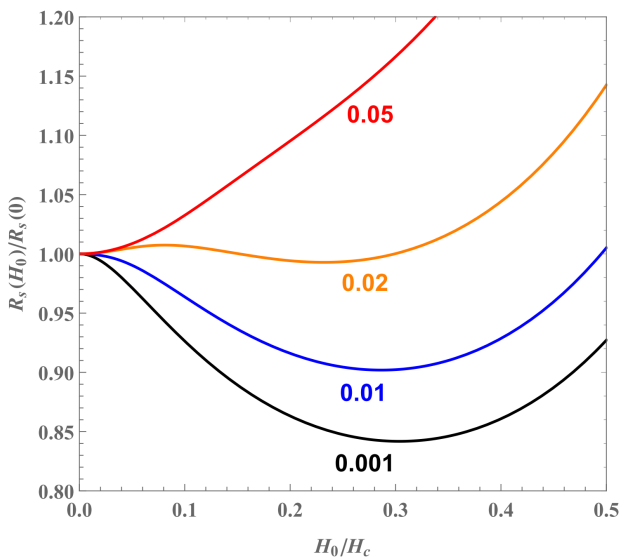}
   \end{center}\vspace{0cm}
   \caption{
$R_s(H_0)$ calculated at $T/\Delta = 0.1$, $\Gamma/\Delta=0.005$, 
$\omega/\Delta= 0.001, 0.01, 0.02, 0.05$. 
   }\label{fig13}
\end{figure}

Shown in Fig.~\ref{fig11} are the Arrhenius plots of $R_s$ versus $1/T$ calculated for different values of $s$ and $\Gamma$. 
At higher temperatures, $\ln R_s$ follows the linear dependence of $1/T$ expected from Eq. (\ref{eq:Rs_MB}) of the BCS model.  
Here moderate broadening of the DOS peaks due to current pairbreaking and subgap states reduce $R_s$.  
At lower temperatures, $\ln R_s$ deviates from the linear dependence and levels off at a finite residual surface resistance, 
which comes from a finite density of subgap states at $\epsilon=0$~\cite{2017_Gurevich_Kubo, 2017_Gurevich_SUST}. Because  
 $n(\epsilon=0)$ depends on $s$ (see Eq. (\ref{noo})), the residual surface resistance also depends on the rf field.

Figure~\ref{fig12} shows the effect of temperature on the field dependence of $R_s(H_0)$. Here
the minimum in $R_s(H_0)$ becomes shallower and shifts to lower fields as $T$ decreases, similar to that is shown in Fig.~\ref{fig7}. 
At the lowest temperature in Fig. ~\ref{fig12}, the monotonic increase of $R_s(H_0)$ with $H_0$ mostly reflects the field dependence of the residual surface resistance.
The effect of the rf frequency on the field dependence of $R_s$ shown in Fig.~\ref{fig13} is similar to that in Fig.~\ref{fig8} for magnetic impurities.

\section{S$^{\prime}$-I-S surface nanostructuring}\label{section_introduction}
In this section we consider a way of tuning the nonlinear surface resistance by a higher - $T_c$ or $H_c$ superconducting (S$^\prime$) layer separated by a thin dielectric (I) layer from the bulk superconductor (S) as shown in Fig. 1b.  Here the I layer is assummed thick enough to suppress the Josephson coupling of the S and S$^\prime$ layers which remain coupled through the electromagnetic interaction of screening currents.  The use of S$^\prime$-I-S multilayers was proposed to increase the superheating field and the onset of dissipative penetration of vortices in superconductors under strong rf fields \cite{mlag1}.  It has been shown \cite{mlag2,mltk1,mltk2,mlc} that, if the penetration depth $\lambda'$ of the S$^\prime$ layer exceeds $\lambda$ of the S-substrate, the  S$^\prime$-I-S structure can withstand the fields exceeding the superheating fields of both S and S$^\prime$ regions at the optimal thickness of the S$^\prime$ layer $d\sim \lambda'$. This results from a current counterflow \cite{mlag2,mltk1} induced in the S$^\prime$ layer by the S-substrate with $\lambda<\lambda'$.  Here we address manifestations of this counterflow effect in the field-dependence of $R_s(H_0)$ for a S$^\prime$-I-S structure.

\subsection{Density of states} \label{SIS_dos}

The retarded Green's functions and $\Delta(x)$ are calculated from the Usadel equations in the $S$ and $S^\prime$ regions with the   
respective pair-breaking parameters $s(x)$ obtained by solving the London equations with the boundary conditions 
$\partial_x A(-d)=\mu_0 H(t)$ and the continuity of $A(x)$ and $\partial_x A(s)$ at the $S-S'$ interface at $x=0$. Using the solution of the London equation \cite{mlag2,mltk1}, we obtain:   
\begin{gather}
s(x,t) = s_0 e^{-2x/\lambda} \sin ^2 \omega t , 
\label{s1}\\
s'(x,t)=s_0'\left(\frac{\lambda}{\lambda'}\cosh\frac{x}{\lambda'}-\sinh\frac{x}{\lambda'}\right)^2\sin^2\omega t 
\label{s2}
\end{gather}
The amplitudes $s_0$ and $s_0'$ are given by
\begin{gather}
s_0=\frac{\Delta}{\pi}\left(\frac{H_0}{H_c}\right)^2C^2, \qquad s_0'=\frac{\Delta'}{\pi}\left(\frac{H_0}{H_c'}\right)^2C^2,
\label{ss0} \\
C = [ \cosh (d/\lambda') + (\lambda/\lambda') \sinh (d/\lambda') ]^{-1}.
\label{C}
\end{gather}
Here the prime corresponds to the respective parameters of the $S'$ layer, and the factor $C$ describes the field attenuation  
by the $S'$ layer at the $S$-$S'$ interface.

\begin{figure}[tb]
   \begin{center}
   \includegraphics[width=0.95\linewidth]{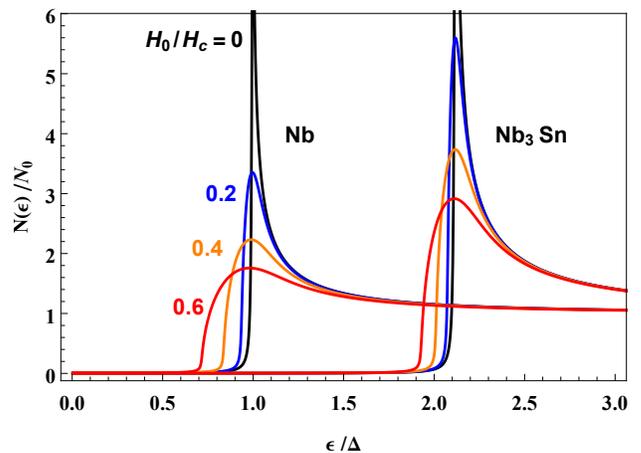}
   \end{center}\vspace{0cm}
   \caption{
The DOS at the ${\rm Nb_3 Sn}$ surface ($x=-d$) and Nb surface ($x=0$) of ${\rm Nb_3 Sn}$-${\rm Nb}$ multilayer calculated for 
$\lambda'/\lambda=3$, 
$d=\lambda'/2$, 
$\sigma_{n}'/\sigma_{n}=0.01$,  
$\Gamma'=\Gamma=0.005$, 
$\Omega'=(4/5)\Omega$, $\Omega=11$, 
$\omega/\Delta=0.001$, and $T/\Delta=0.12$. 
   }\label{fig14}
\end{figure}
Shown in Fig.~\ref{fig14} is the DOS calculated for a ${\rm Nb_3 Sn}$ layer on the bulk ${\rm Nb}$. 
Here the DOS in the $S'$ layer is less broadened as compared to the $S$ region, 
even though it is the $S'$ layer which is directly exposed to the external field. Such reduction of the pairbreaking effect 
in the $S'$ layer results from a counterflow induced by the $S$ substrate  if $\lambda' > \lambda$~\cite{mlag2,mltk1}. 
This condition is satisfied for Nb$_3$Sn-Nb multilayers with typical values of $\lambda'\simeq (2-3)\lambda$.

\subsection{Surface resistance} \label{section:Rs}

The surface resistance of the $S'$-$I$-$S$ structure is:  
\begin{gather}
R_{s}  = \frac{\mu_0^2\omega^2\lambda^3\Delta C^2 }{2\rho_n T } \biggl[ \frac{1}{s_{0}}\int_0^{s_{0}} I(s,\omega)ds + 
\nonumber \\
 \frac{2 \lambda'^2}{\lambda^3} \frac{\sigma_s'}{\sigma_s} \int_{-d}^{0}I'[s'(x)]\left(\frac{\lambda}{\lambda'}\cosh\frac{x}{\lambda'}-\sinh\frac{x}{\lambda'}\right)^2\!dx\biggr]. 
\label{eq:Rs_b}
\end{gather}
Here the first term in the brackets describes the contribution of the $S$ region in the same way as in Eq. (\ref{RsG}), where $I(s,\omega)$ is defined by Eq. (\ref{I}), 
and $C^2$ accounts for the field attenuation by the $S'$ layer. If $S$ and $S'$ are made of the same material,  Eq.~(\ref{eq:Rs_b}) reduces to Eq. (\ref{RsG}).

\begin{figure}[t]
   \begin{center}
   \includegraphics[width=0.95\linewidth]{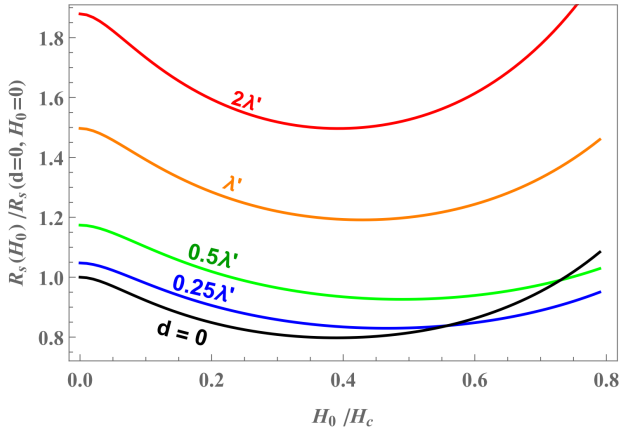}
   \end{center}\vspace{0cm}
   \caption{
$R_s(H_0)$ of a Nb-Nb bilayer calculated for 
$d/\lambda' = 0, 0.25, 0.5, 1, 2$, 
$\lambda'/\lambda=2$, 
$\sigma_{n}'/\sigma_{n}=0.25$,  $\Omega'=\Omega=11$, 
$\Gamma'=\Gamma=0.005$, 
$\omega/\Delta=0.001$, $T/\Delta=0.12$. 
   }\label{fig15}
\end{figure}
We first use Eq. (\ref{eq:Rs_b}) to calculate $R_s(H_0)$ for a dirtier $S'$ layer deposited onto the same but a cleaner material $S$ with $\Delta=\Delta'$. 
For instance,
Fig.~\ref{fig15} shows the results for a dirty Nb layer on a cleaner Nb substrate. 
Here the dirtier layer always increases $R_s(H_0)$ at low fields consistent with the Mattis-Bardeen Eq. (\ref{eq:Rs_MB}). However,  
as $H_0$ increases $R_s(H_0)$ at $d\sim (0.25-0.5)\lambda'$ becomes smaller than $R_s(H_0)$ at $d=0$.  
Mitigation of the high-field increase of $R_s(H_0)$ by a dirty surface layer reflects the counterflow effect which reduces the surface current in the $S'$ layer, while the $S'$ layer partially screens the bulk $S$ region. As a result, the minimum in $R_s(H_0)$ shifts to a higher field at $d\simeq d_m$, where $d_m$ is an optimal thickness  at which the $S$-$I$-$S'$ structure can screen the magnetic field exceeding the bulk superheating fields of both $S$ and $S'$ ~ \cite{mlag2,mltk1}.
The minimum in $R_s(H_0)$ moves back to $H_0 \sim 0.4 H_c$ if $d \gtrsim 2\lambda'$. 

\begin{figure}[t]
   \begin{center}
   \includegraphics[width=0.95\linewidth]{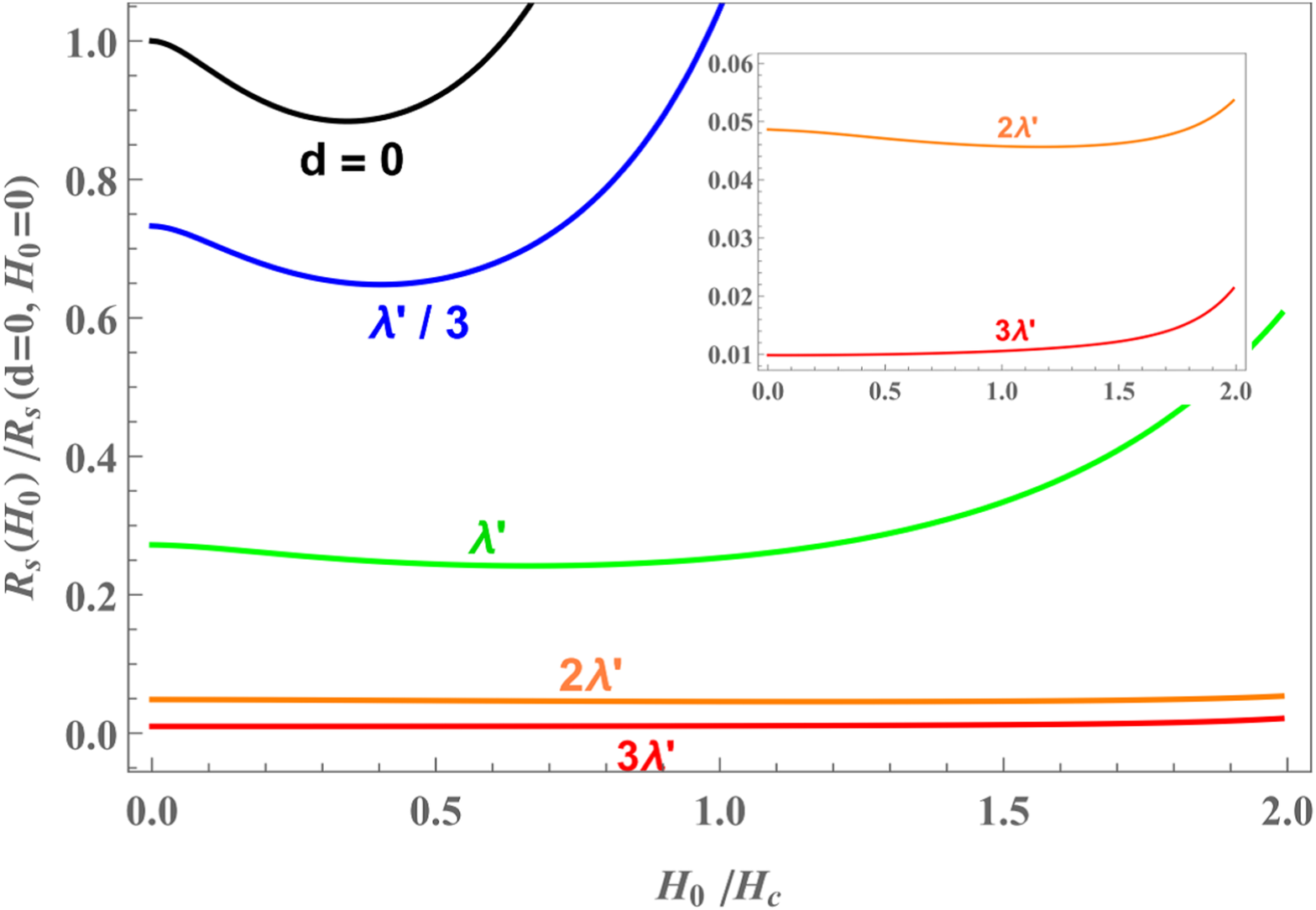}
   \end{center}\vspace{0cm}
   \caption{
$R_s(H_0)$ of a ${\rm Nb_3 Sn}$-Nb bilayer calculated for 
$d/\lambda'=0, 1/3, 1, 2, 3$, 
$\lambda'/\lambda=3$, 
$\sigma_{n}'/\sigma_{n}=0.01$, 
$\Omega'=(4/5) \Omega$, $\Omega'=11$, 
$\Gamma'=0.05$, $\Gamma=0.01$, 
$\omega/\Delta=0.001$, and $T/\Delta=0.12$. 
   }\label{fig16}
\end{figure}

Next we calculate $R_s(H_0)$ for a $S'$-$I$-$S$ structure made of different materials with $\Delta'>\Delta$ in which case the behavior of $R_s(H_0)$ changes significantly as compared to the above case of $\Delta=\Delta'$. 
For instance, Fig.~\ref{fig16} shows $R_s(H_0)$ calculated for a ${\rm Nb_3 Sn}$-${\rm Nb}$ multilayer. 
Because $\Delta'_{\rm{Nb_3Sn}}\simeq 2\Delta_{\rm{Nb}}$, the surface resistance of Nb$_3$Sn is smaller than $R_s$ for Nb at $T\ll T_c$. As a result, 
$R_s$ reduces as $d$ increases, in contrast to the case for a Nb-Nb multilayer. However,
thick ${\rm Nb_3 Sn}$ films with $d \gg \lambda'$ are prone to the dissipative vortex penetration at lower fields $H_0 \sim H_{c1}^{\rm{Nb_3Sn}}\ll H_{c1}^{\rm{Nb}}$, 
resulting in a significant increases of $R_s$. Thus, we are interested in optimum thicknesses $d \lesssim \lambda'$ for which the field onset of vortex 
penetration is greatly increased \cite{mlag1} while the screening field limit is pushed up beyond the superheating fields of both $S$ and $S'$ superconductors \cite{mlag2,mltk1}.
As an illustration, Fig.~\ref{fig16} shows $R_s(H_0)$ calculated for $\Gamma'=0.05\Delta$ and $\Gamma=0.01\Delta$.  
In this case the minimum in $R_s(H_0)$ in a thick $S'$ layer with $d\gg \lambda'$ disappears due to the large $\Gamma'$. However,
a thin $S'$ coating with $d \sim \lambda'$ not only preserves the minimum in $R_s(H_0)$ but also pushes it to higher fields exceeding $H_c$ of the $S$ substate. 
These results suggest that a thin $\rm{Nb_3Sn}$ coating can be used to produce high gradient resonant cavities with no field degradation of the quality factors $Q(H_0)$ 
which can exceed $Q^{\rm{Nb}}(H_0)$ up to fields close to the superheating field of ${\rm Nb_3 Sn}$.

\section{Proximity coupled normal layer at the surface}

In this section we consider the effect of a thin N layer on $R_s(H_0)$, generalizing our approach developed for the calculations of $R_s$ at low fields \cite{2017_Gurevich_Kubo} to the case of strong current pairbreaking in type-II superconductors with $\lambda\gg\xi$ and proximity-coupled systems \cite{1970_Usadel, 1999_Belzig_review, 2004_Golubov_review}.  For the geometry shown in Fig.~\ref{fig1}(b), $\theta(x)$ is nearly constant across a thin N layer with $d_N\ll\xi_N$, so the problem can be reduced to solving the Usadel equation only in the S region and taking into account the effect of the N layer by a self-consistent boundary condition at $x=0$. In what follows we outline 
the essential steps of this approach and give the final formulas necessary for the calculations of $R_s(H_0)$, relegating all technical details to Appendix \ref{appendix:Proxi}.  

\subsubsection{Retarded Green's functions and density of states}

The real-frequency Usadel equation for the functions $u$ and $v$ in the S-region can be written in the form
\begin{gather}
u'(x)^{2} - v'(x)^{2} = g_1(u,v) - g_1(u_s,v_s)  , 
\label{x-dependence1}\\
2 u'(x) v'(x) = g_2(u,v) - g_2(u_s,v_s) , 
\label{x-dependence2}
\end{gather}
where
\begin{gather}
g_1(u,v) = \frac{s}{2} \cosh 2u \cos 2v + 2 ( \Gamma_{\rm S} \cosh u \cos v  \nonumber \\
+ \epsilon \sinh u \sin v   - \Delta_s \cosh u \sin v ) 
\label{g1}\\
g_2(u,v) = \frac{s}{2} \sinh 2u \sin 2v + 2 (\Gamma_{\rm S} \sinh u \sin v \nonumber \\
-\epsilon \cosh u \cos v + \Delta_s \sinh u \cos v ). 
\label{g2}
\end{gather}
Here $u_s$ and $v_s$ are the solutions of the uniform Usadel Eq. (\ref{eq:R_Usadel_Gamma_1}) and (\ref{eq:R_Usadel_Gamma_2}).
As shown in Appendix \ref{appendix:Proxi}, Eqs. (\ref{x-dependence1}) and (\ref{x-dependence2}) are supplemented by the following boundary condition on the S-side  
at $x=+0$:
\begin{gather}
 u_0' = \alpha \biggl( \frac{s}{2r}\sinh 2u_{-}\cos 2v_{-} + \epsilon \cosh u_{-}\sin v_{-} \nonumber \\
+ \Gamma_{\rm N} \sinh u_{-}\cos v_{-}  \biggr) + \Psi \sinh u_0 \sin v_0  , 
\label{bc2re} \\
 v_0' = \alpha \biggl( \frac{s}{2r}\cosh 2u_{-}\sin 2v_{-} - \epsilon \sinh u_{-}\cos v_{-} \nonumber \\
+ \Gamma_{\rm N} \cosh u_{-}\sin v_{-}  \biggr) - \Psi \cosh u_0 \cos v_0. 
\label{bc2im}
\end{gather}
According to the boundary condition (\ref{eq:BC1}), the complex function $\theta(x)=u(x)+iv(x)$ has a jump at $x=0$ for any nonzero contact resistance $R_B$. Here $\theta(-0)=u_{-}+iv_{-}$ on the N side of the N-S interface is related to $\theta(x+0)=u_0+iv_0$ by two coupled equations:
\begin{gather}
\beta \biggl[ \frac{s}{2r}\sinh 2u_{-}\cos 2v_{-} + \epsilon \cosh u_{-}\sin v_{-} + \nonumber \\
\Gamma_{\rm N} \sinh u_{-}\cos v_{-}  \biggr] = \sinh (u_0 - u_{-}) \cos (v_0 -v_{-} ) , 
\label{bc1re}\\
\beta \biggl[ \frac{s}{2r}\cosh 2u_{-}\sin 2v_{-} - \epsilon \sinh u_{-}\cos v_{-} +\nonumber \\
\Gamma_{\rm N} \cosh u_{-}\sin v_{-}  \biggr] = \cosh (u_0 - u_{-}) \sin (v_0 -v_{-} ), 
\label{bc1im}
\end{gather}
where $r=D_S/D_N$, and the parameter $\Psi$ in Eq. (\ref{bc2re}) and (\ref{bc2im}) which accounts for a short-range disturbance of the pair potential caused by the N layer in the S region was calculated in Appendix \ref{appendix:Proxi}:
\begin{gather}
\Psi = \frac{2\pi T g\alpha\eta \sum_{\omega>0} k_\omega^{-2}\cos\theta_s }{1-2\pi T g \sum_{\omega>0} k_\omega^{-2}\cos^2\theta_s}, 
\label{eq:Psi} \\
k_\omega^2 = s\cos 2\theta_s + (\omega_n + \Gamma_{\rm S})\cos\theta_s + \Delta \sin\theta_s.
\label{k2} 
\end{gather}
Here $\Delta(x)$ is given by Eq. (\ref{Deltsg}), and $\theta_s$ is determined by the uniform thermodynamic Usadel equation:
\begin{equation} 
(\omega_n+\Gamma_s)\sin\theta_s+s\sin\theta_s\cos\theta_s=\Delta\cos\theta_s
\label{usau}
\end{equation}
\begin{figure}[tb]
   \begin{center}
   \includegraphics[width=0.95\linewidth]{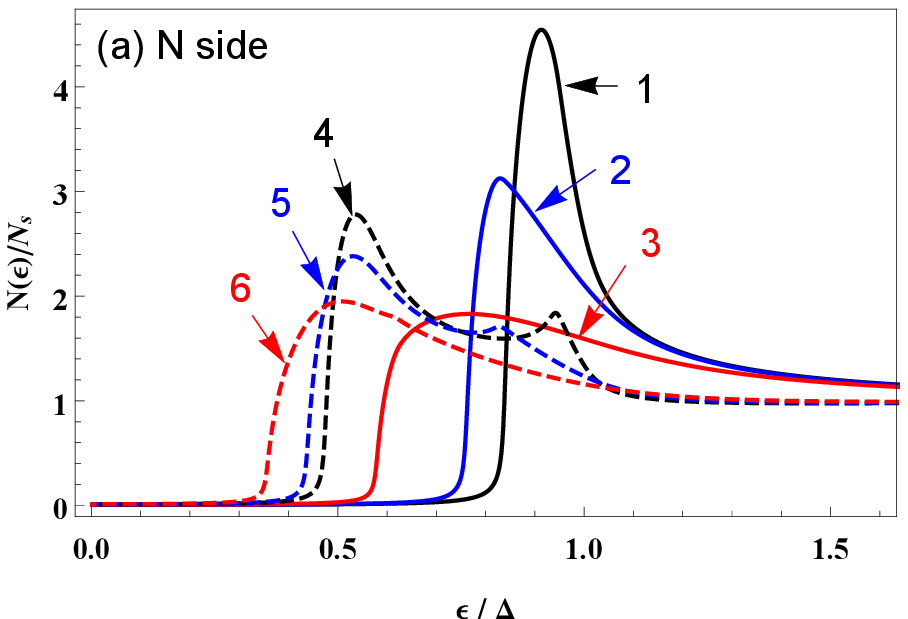}
   \includegraphics[width=0.95\linewidth]{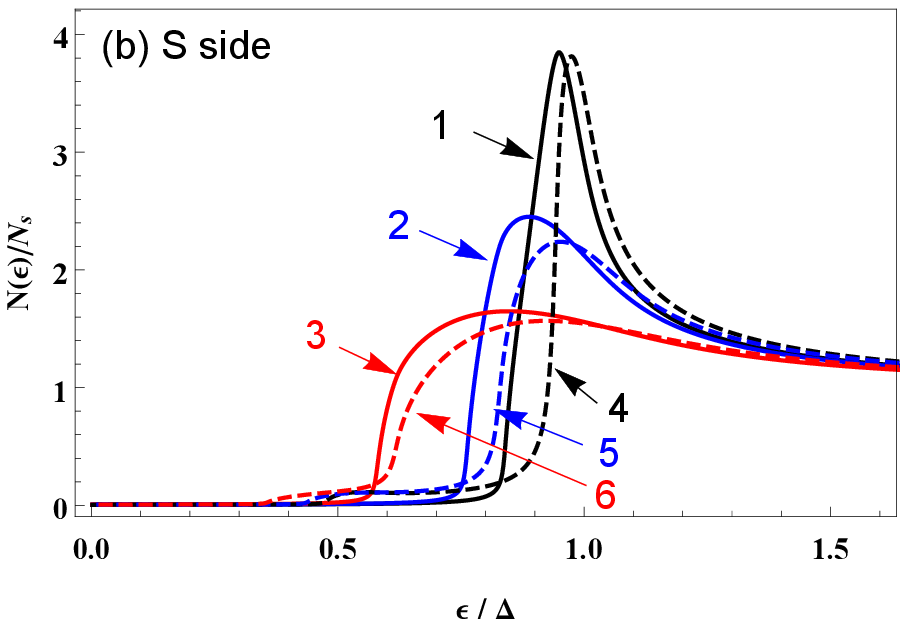}
   \end{center}\vspace{0cm}
   \caption{
Densities of states at (a) N side and (b) S side of the interface calculated for $\alpha=0.05$, $\Gamma_{\rm N}/\Delta= \Gamma_{\rm S}/\Delta = 0.005$, $\Omega /\Delta = 11$, 
$(\beta, \, s)=(0.1, \, 0.005)$ for 1, $(0.1, \, 0.03)$ for 2, $(0.1, \, 0.1)$ for 3, 
$(1, \, 0.005)$ for 4, $(1, \, 0.03)$ for 5, and $(1, \, 0.1)$ for 6.  
   }\label{fig17}
\end{figure}
\begin{figure}[tb]
   \begin{center}
   \includegraphics[width=0.95\linewidth]{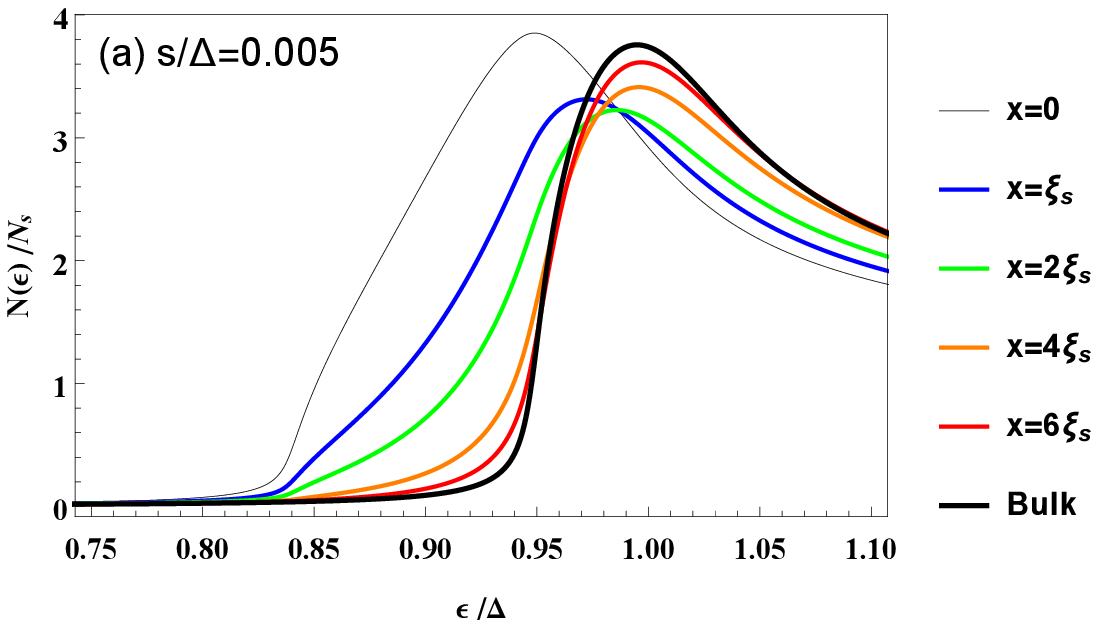}
   \includegraphics[width=0.95\linewidth]{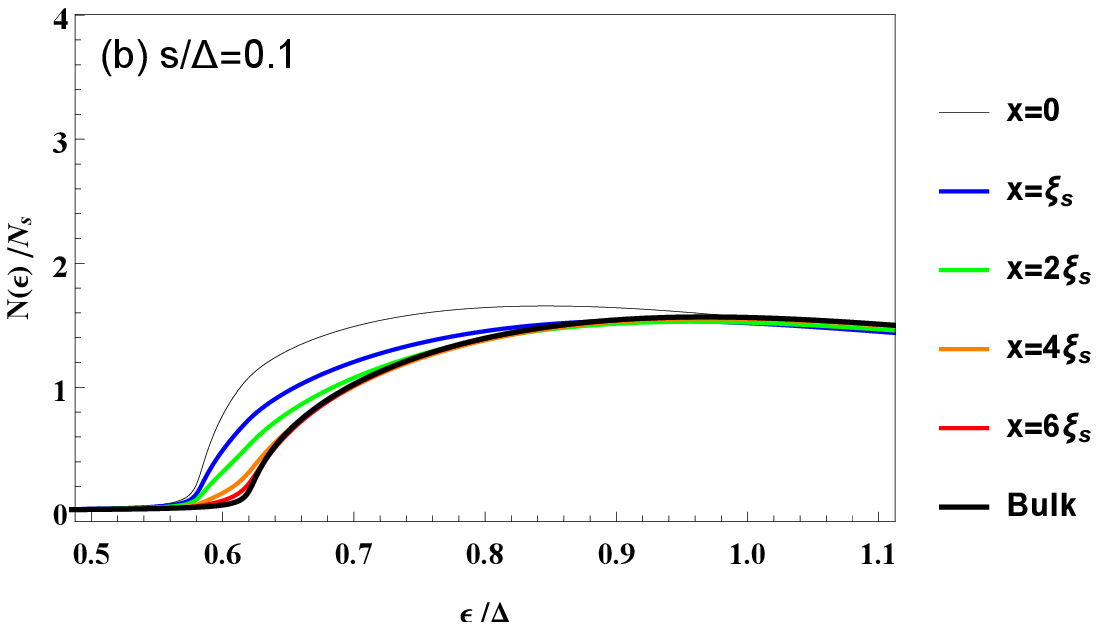}
   \end{center}\vspace{0cm}
   \caption{
DOS in the S region at different depths: $x/\xi_s = 0, 1, 2, 4, 6$, and $s/\Delta=$ (a) 0.005 and (b) 0.1. 
Here $\alpha=0.05$, $\beta=0.1$, 
$\Gamma_{\rm N}/\Delta= \Gamma_{\rm S} /\Delta= 0.005$, and
$\Omega /\Delta= 11$. 
   }\label{fig18}
\end{figure}

The closed set of Eqs.~(\ref{x-dependence1})-(\ref{usau}) determine self-consistently the Green functions and the pair potential. We solved these equations numerically 
to calculate the spectral function $M(\epsilon,s)$ and the DOS which are the key input characteristics in the calculations of $R_s(H_0)$.   
For instance, Fig.~\ref{fig17} shows the DOS at (a) the N and (b) S side of the interface at $x=0$. 
For a small contact resistance $R_B$, that is, a nearly transparent interface with $\beta \ll 1$ (solid curves), 
the N and S regions are strongly coupled and the behaviors of both $N_N(\epsilon)$ and $N_S(\epsilon)$ are similar to that of $N(\epsilon)$ shown in Fig. \ref{fig9} for 
an ideal surface. As $R_B$ and $\beta$ increases (dashed curves), 
the proximity coupling of the N and S regions weakens, and minigap states in the N layer appear. These states are further broaden and shifted to lower energies by current. 

Figure~\ref{fig18} shows the evolution of the DOS in the S region with the distance from the NS interface calculated for: (a) $s/\Delta=0.005$ and (b) $s/\Delta=0.1$. 
As $x$ increases, the effect of the N layer on the DOS weakens and $N(\epsilon)$ approaches the bulk DOS in the S region.  
A spatial extent of the DOS disturbance by the proximity effect is given by  $L \sim \xi_s (\Delta/\delta \epsilon)^{1/4}$, 
where $\delta \epsilon$ is the DOS peak width ~\cite{2017_Gurevich_Kubo}. As $s$ increases $\delta \epsilon$ increases and the proximity induced DOS disturbance becomes more short-ranged.

\subsubsection{Surface resistance}

%
\begin{figure}[tb]
   \begin{center}
   \includegraphics[width=0.95\linewidth]{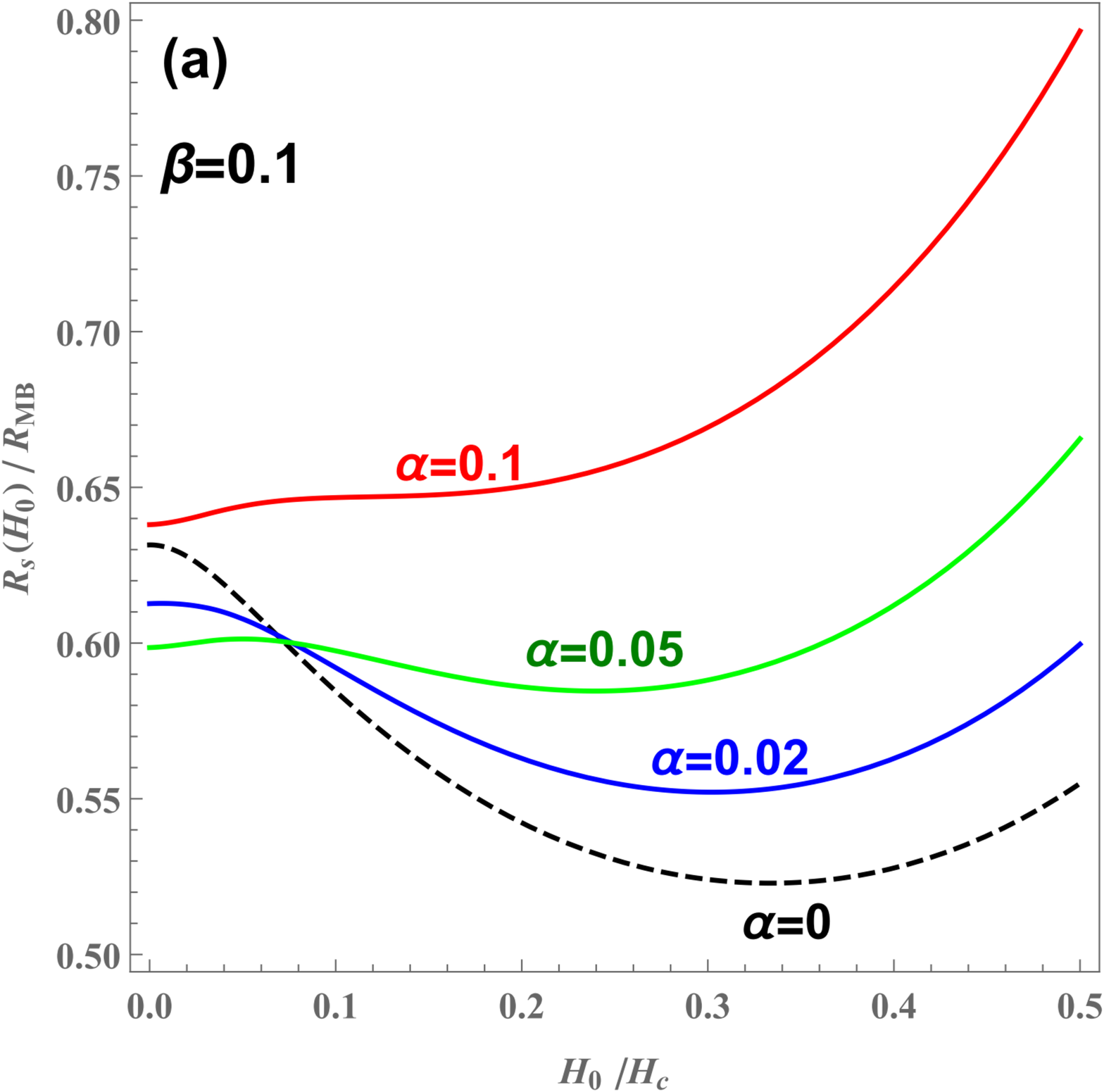}
   \includegraphics[width=0.95\linewidth]{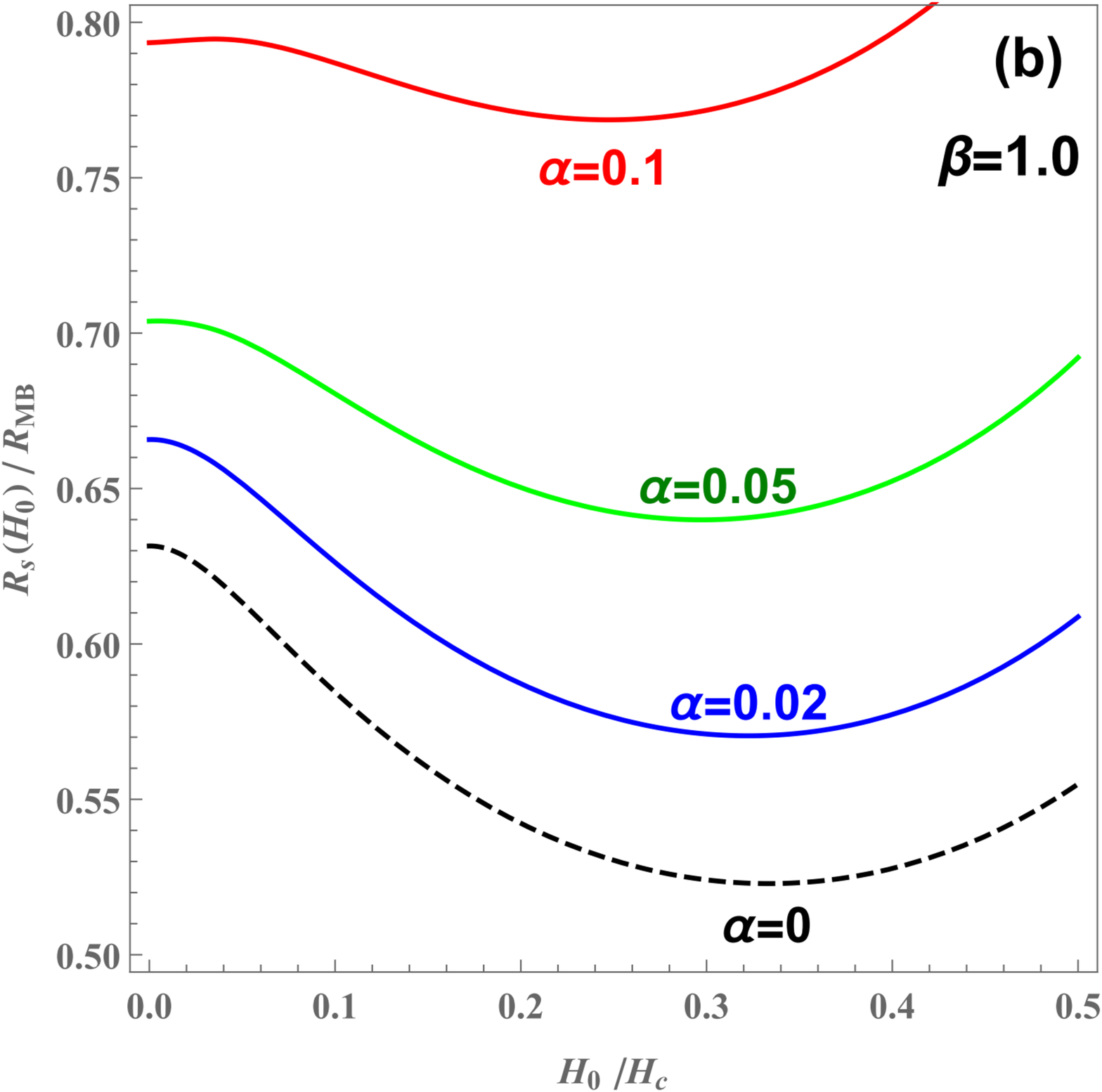}
   \end{center}\vspace{0cm}
   \caption{
$R_s(H_0)$ calculated at different thicknesses of the N layer:  $\alpha=0.02, 0.05, 0.1, 0.2$, 
$\beta=0.1, 1.0$, $\Gamma_{\rm N}/\Delta= \Gamma_{\rm S}/\Delta = 0.005$, $\Gamma_p=0$, 
$\Omega /\Delta = 11$, $D_n=0.5 D_s$, $\lambda=10\xi_s$, 
$\omega/\Delta = 0.001$, and $T/\Delta = 0.11$. 
The dashed line shows $R_s(H_0)$ calculated for $d=0$, $\Gamma/\Delta=0.005$ and $\Gamma_p=0$. All $R_s(H_0)$ curves are normalized to 
the ideal BCS surface resistance $R_{MB}$ at $\Gamma=0$.
   }\label{fig19}
\end{figure}
\begin{figure}[tb]
   \begin{center}
   \includegraphics[width=0.95\linewidth]{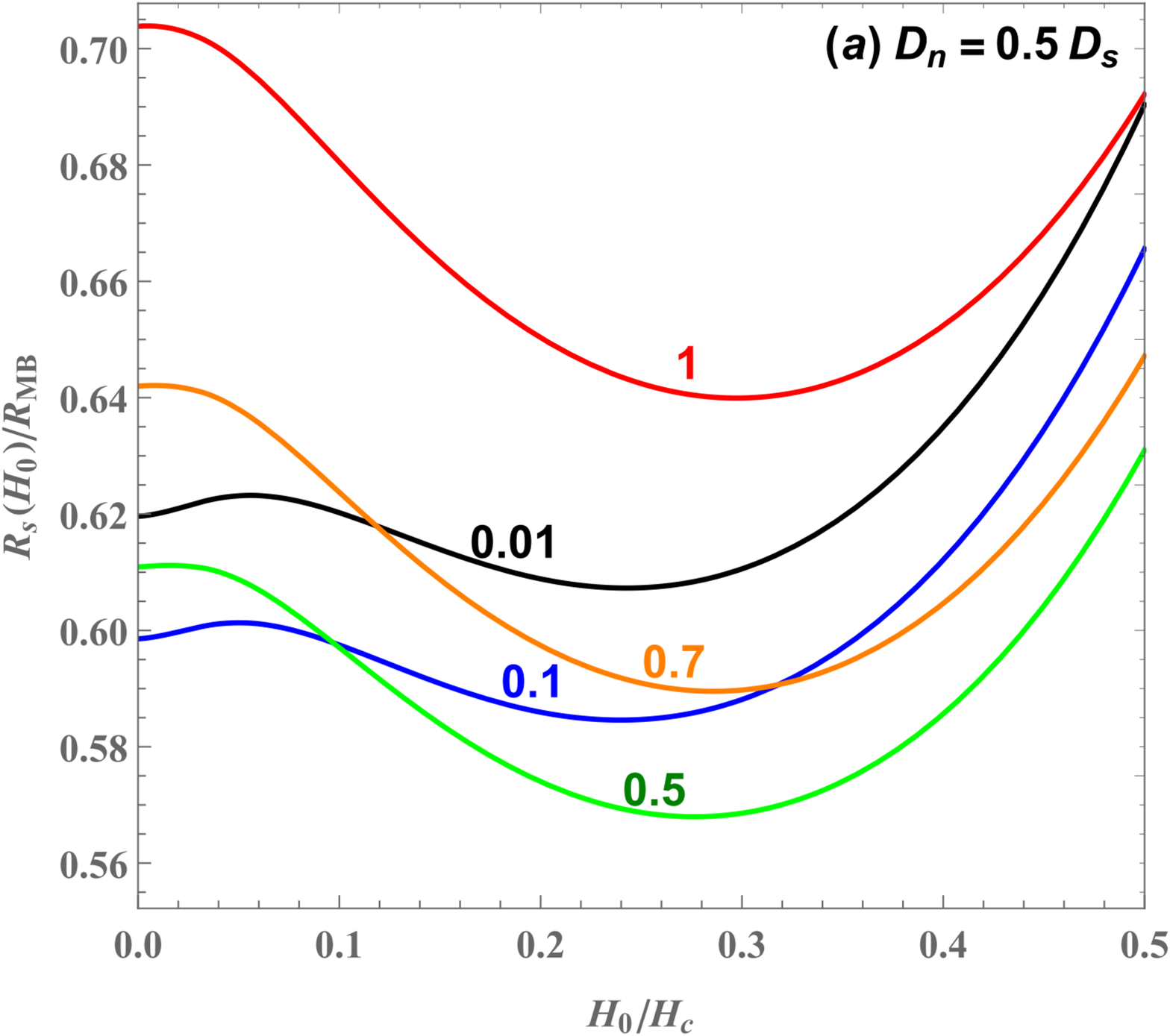}
   \includegraphics[width=0.95\linewidth]{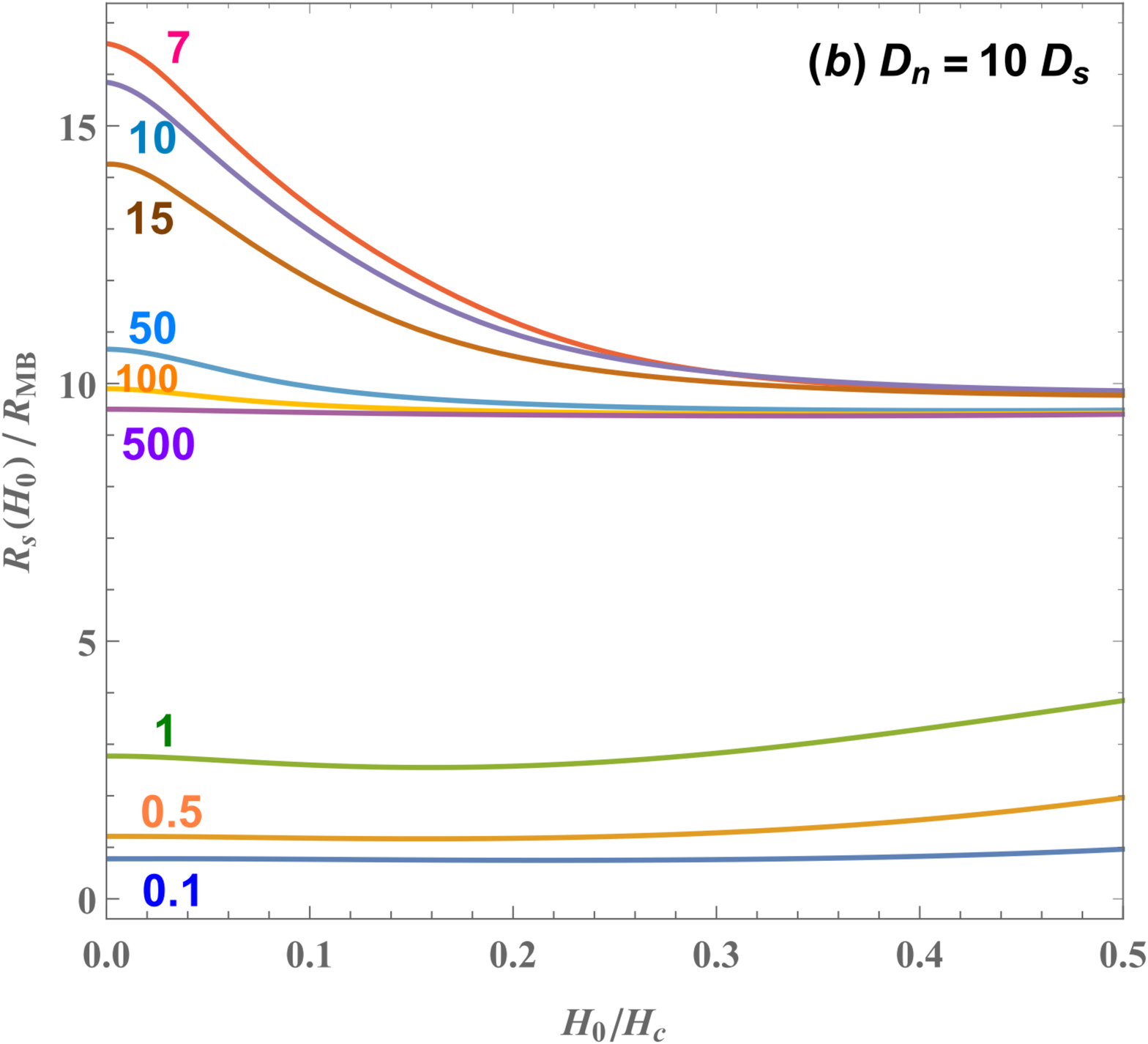}
   \end{center}\vspace{0cm}
   \caption{
$R_s(H_0)$ calculated at  
(a) $D_n=0.5D_s$, $\beta=0.01, 0.1, 0.5, 0.7, 1$, 
(b) $D_n=10D_s$, $\beta=0.1, 0.5, 1, 7, 10, 15, 50, 100, 500$, 
$\alpha=0.05$, 
$\Gamma_{\rm N}/\Delta= \Gamma_{\rm S}/\Delta = 0.005$, $\Gamma_p=0$,
$\Omega /\Delta = 11$,  $D_n/D_s=0.5, 10$, $\lambda=10\xi_s$, 
$\omega/\Delta = 0.001$, and $T/\Delta = 0.11$. 
   }\label{fig20}
\end{figure}
\begin{figure}[tb]
   \begin{center}
   \includegraphics[width=0.95\linewidth]{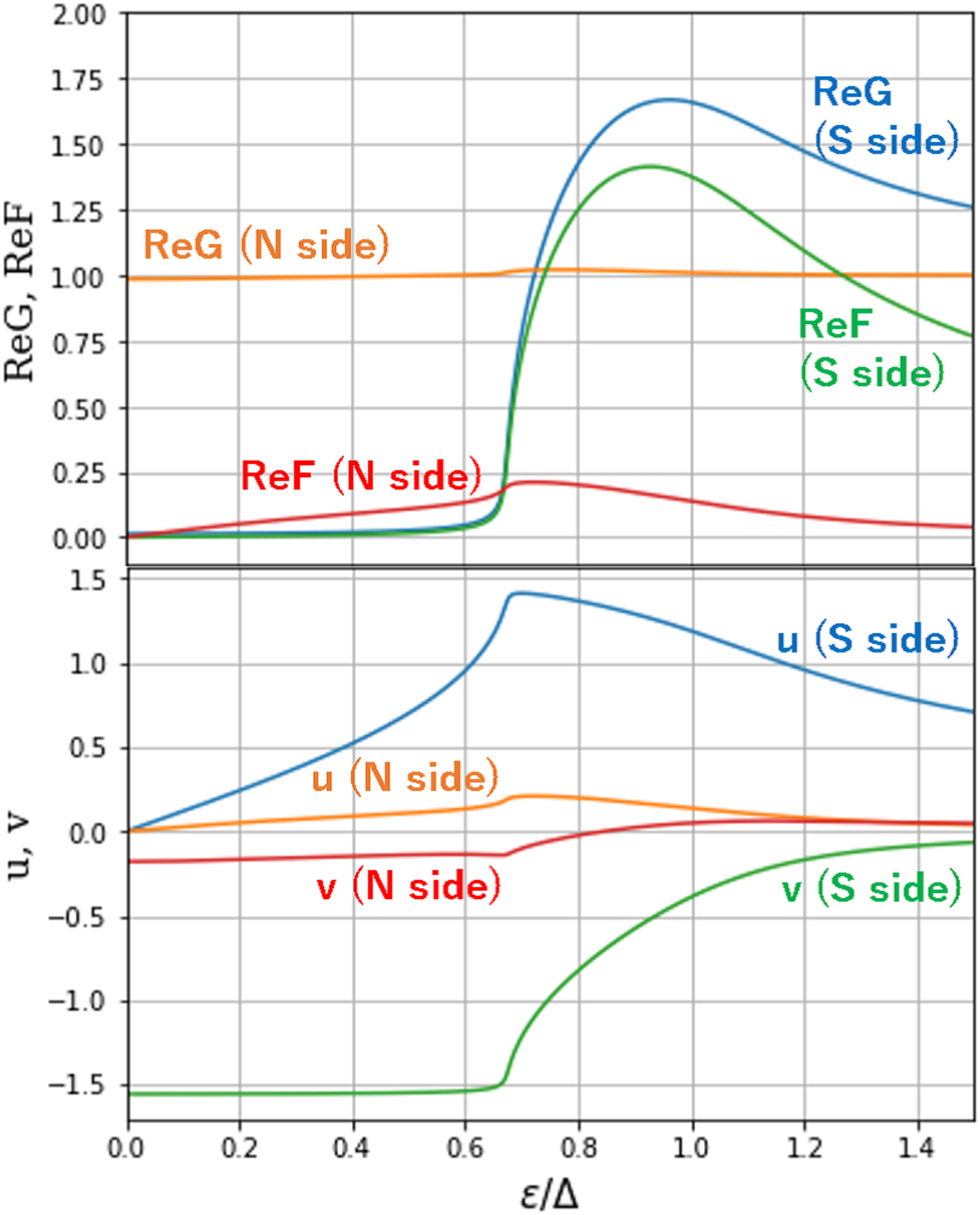}
   \end{center}\vspace{0cm}
   \caption{
Green's functions at the NS interface calculated at 
$H_0=0.5H_c$, 
$\alpha=0.05$, $\beta=7$, 
$\Gamma_{\rm N}= \Gamma_{\rm S} = 0.005\Delta$, $\Gamma_p=0$,  
$\hbar \Omega/\Delta = 11$, $\lambda=10\xi_s$, $D_n = 10 D_s$, 
$\hbar \omega /\Delta= 0.001$, and $k_B T /\Delta = 0.11$. 
   }\label{fig21}
\end{figure}
\begin{figure}[tb]
   \begin{center}
   \includegraphics[width=0.95\linewidth]{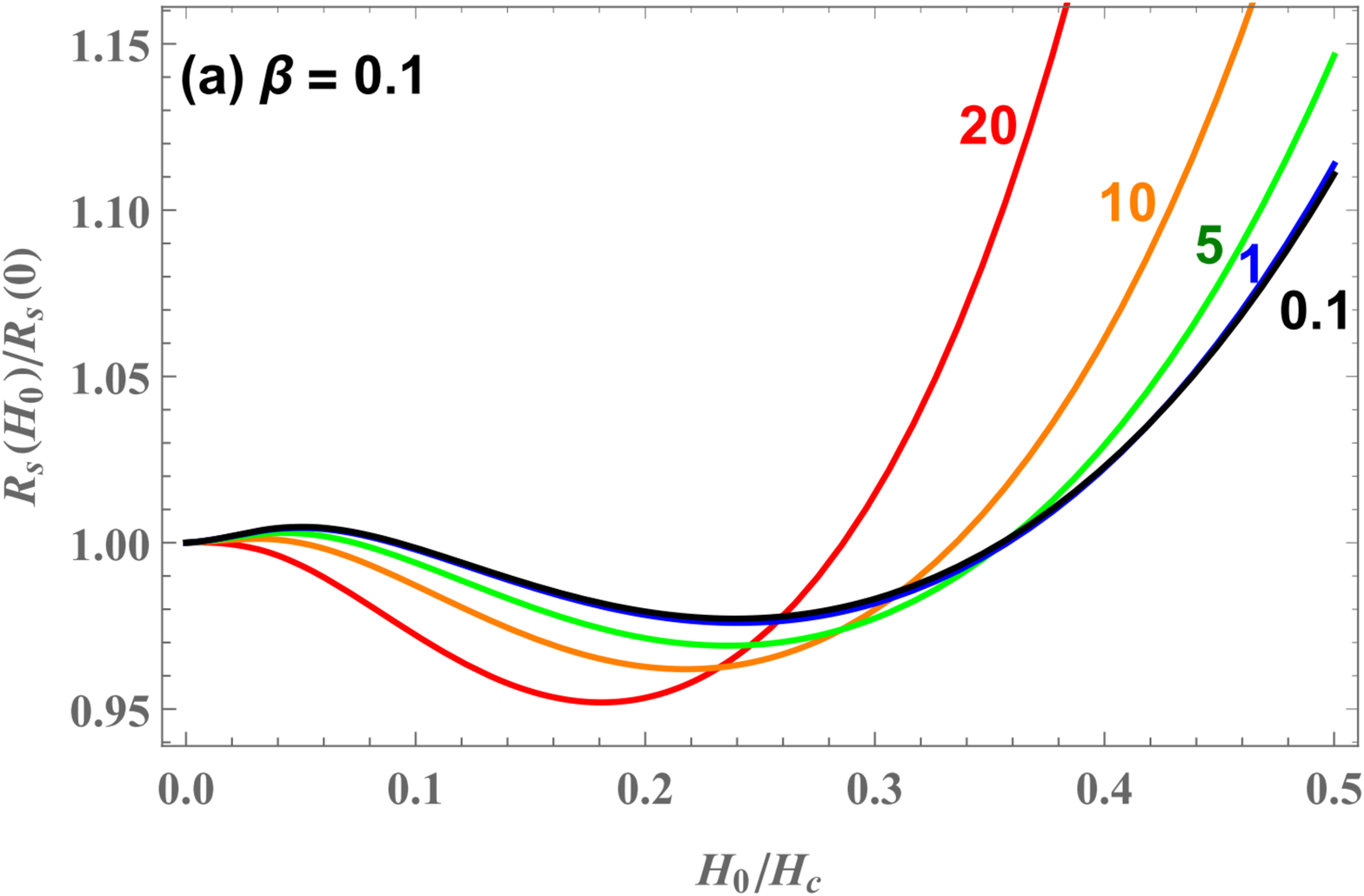}
   \includegraphics[width=0.95\linewidth]{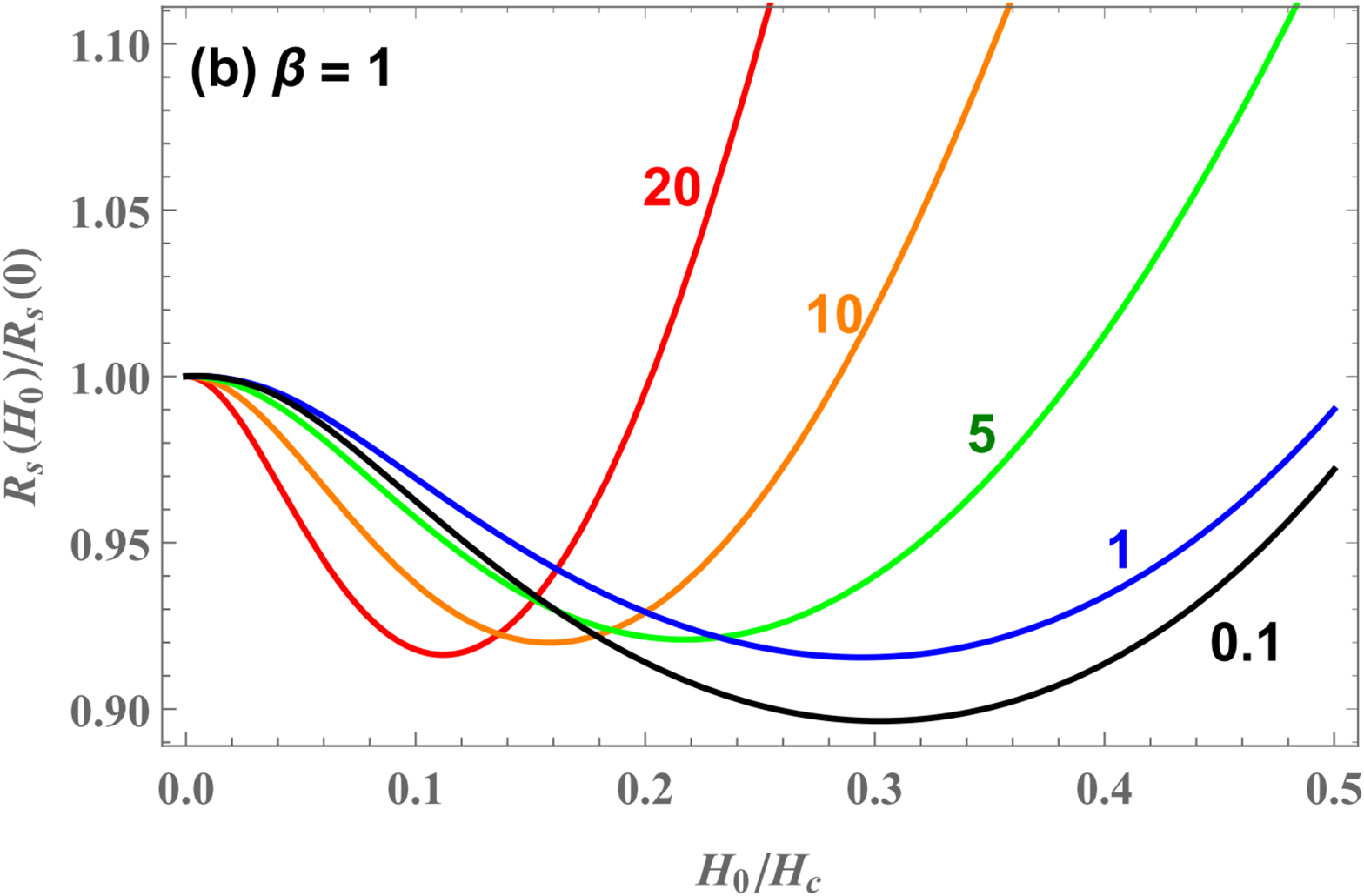}
   \includegraphics[width=0.95\linewidth]{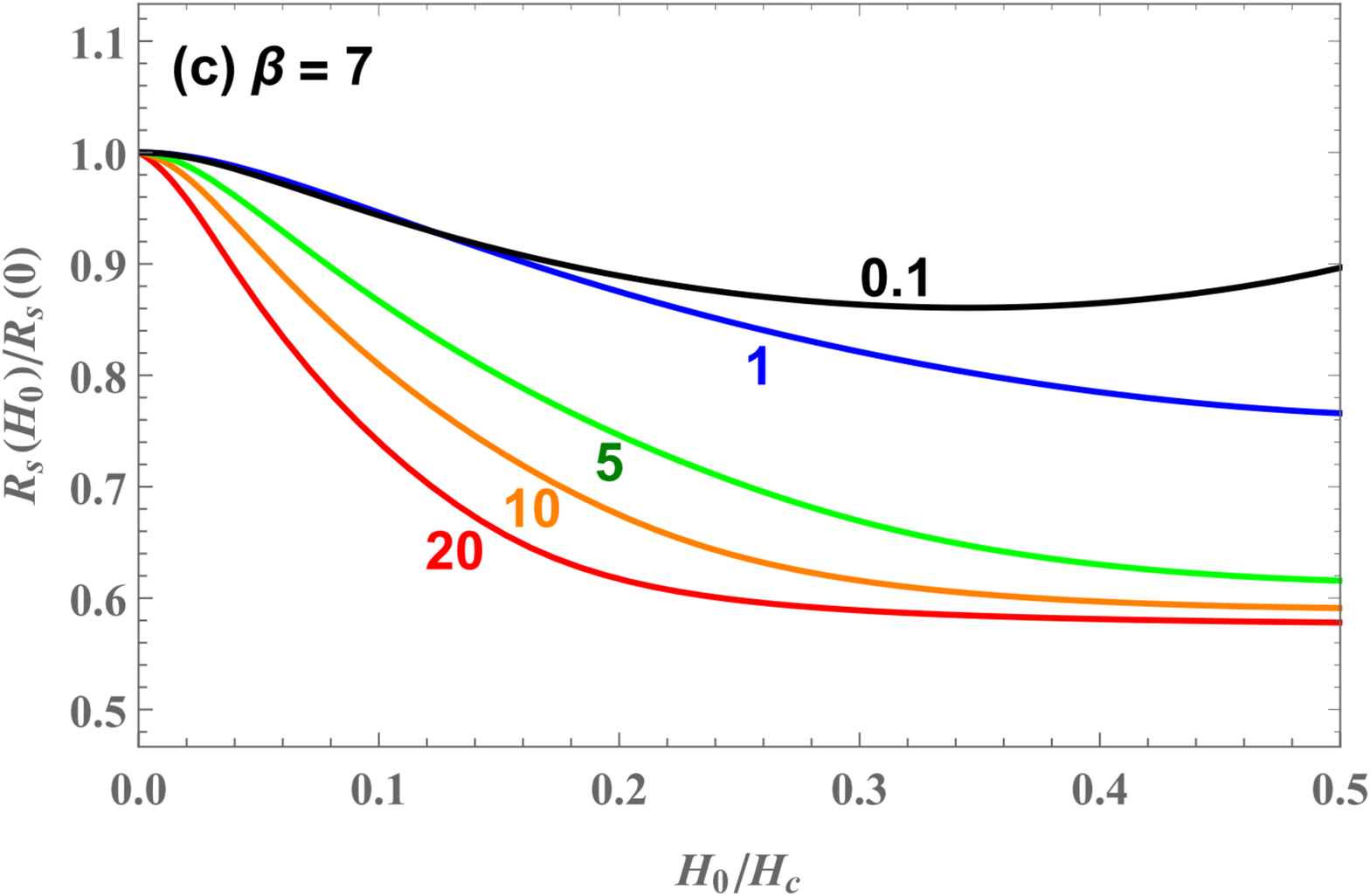}
   \end{center}\vspace{0cm}
   \caption{
$R_s(H_0)$ calculated for $D_n/D_s = 0.1, 1, 5, 10, 20$, 
$\alpha=0.05$, $\beta=0.1, 1, 7$, $\Gamma_{\rm N}= \Gamma_{\rm S} = 0.005\Delta$, $\Gamma_p=0$,  
$\hbar \Omega/\Delta = 11$, $\lambda=10\xi_s$, 
$\hbar \omega /\Delta= 0.001$, and $k_B T /\Delta = 0.11$. 
   }\label{fig22}
\end{figure}

The surface resistance is given by Eqs.~(\ref{eq:Rs_b}):
\begin{equation}
\!R_s = 
2\mu_0^2\omega^2\lambda^2\frac{\Delta}{T}\left[ 
d\sigma_n I(0)
+ \sigma_s\!\!\int_0^{\infty}\!\!I(x)e^{-2x/\lambda}dx\right]\!\!.
\label{rsn}
\end{equation}
Here the first term in the brackets comes from the contribution of the N layer. The integral term describes the contribution of the S region, where $I(x)$ defined by Eq. (\ref{I})  
is calculated by solving Eqs.~(\ref{x-dependence1})-(\ref{usau}) taking into account both the proximity effect and the spatial variation of the current pairbreaking parameter $s(x)$.  

Shown in Fig.~\ref{fig19} is $R_s (H_0)$ calculated  from Eq. (\ref{rsn}) for different thicknesses of the N layer, where 
the dashed curves correspond to $R_s(H_0)$ for an ideal surface with $\Gamma=0.005\Delta$. In all cases  
the minimum in $R_s (H_0)$ is due to the interplay of the current-induced DOS broadening and the reduction of the quasiparticle gap. 
As a result, the N layer shifts the minimum in $R_s(H_0)$ to lower fields as $d$ increases 
because the proximity-induced DOS broadening becomes more pronounced as the N layer gets thicker, so that 
the optimum width of DOS peaks is achieved at smaller $H_0$. 
For strongly-coupled N layers represented in Fig.~\ref{fig19}(a), the minimum in $R_s(H_0)$ 
disappears as the N layer thickness exceeds a critical value $d>d_c=\alpha \xi_sN_s/N_n$ which is still much smaller than $\xi_s$ at $\alpha\simeq 0.1$. 

As follows from Fig.~\ref{fig19}(a), for thin strongly-coupled N layers with $\alpha\lesssim 0.05$, there is a crossover in the field dependence of $R_s(H_0)$: the low-field $R_s(H_0)$ is smaller than the corresponding $R_s$ for an ideal surface but this relation reverses as $H_0$ increases. This reduction of $R_s$ at small $d$ comes from the long range proximity effect in which the length $L \sim \xi_s (\Delta/\delta \epsilon)^{1/4}$ of the DOS disturbance produced by the N layer extends deep into the S region providing an optimum DOS broadening ~\cite{2017_Gurevich_Kubo}. The resulting reduction of $R_s (H_0=0)$ in the S region overwhelms the increase of $R_s$ due to dissipation in the N layer if  $\delta \epsilon \sim \Gamma \ll \Delta$. However,  as $H_0$ increases, the current-induced DOS broadening takes over, shortening the range of DOS disturbance $L$ as shown in Fig.~\ref{fig18} and increasing $R_s$ in the S region. 
It should be noted that this behavior of  $R_s(H_0)$ is only characteristic of strongly-coupled N layers with $\beta\ll 1$. For weakly coupled N layers with large contact resistances and $\beta > 1$, the field crossover in $R_s(H_0)$ disappears, as shown in  Fig.~\ref{fig19}(b).

The effects of contact resistance on $R_s(H_0)$ is shown in Fig.~\ref{fig20}. 
In the case of $D_n=0.5D_s$ represented by Fig.~\ref{fig20}(a) the proximity effect reduces $R_s$ at low fields at $\beta \lesssim 0.1$ but the optimal $\beta$  
at high RF fields becomes much larger. 
For instance, the deepest minimum in $R_s(H_0)$ and the lowest high field $R_s(H_0)$ occur at intermediate coupling $\beta\simeq 0.5$. 
For larger $\beta\gtrsim 1$, the minigap reduction in the N layer becomes dominant mechanism which increases $R_s(H_0)$.  
A rather different evolution $R_s(H_0)$ with $\beta$ occurs for a highly conductive N layer with $D_n=10D_s$ represented in Fig. ~\ref{fig20}(b). 
In this case the N layer gives the main contribution to $R_s$ but the effect of the contact resistance on $R_s$ is rather intricate: $R_s$ first increases with $\beta$ reaching the maximum low-field values at $\beta\simeq 7$ and then decreases as $\beta$ further increases. Here $R_s(H_0))$ at $\beta \gtrsim 3$ {\it exceeds} $R_s$ of a completely decoupled N layer at $\beta\to\infty$. Such unexpected behavior can be understood by analysing the Green functions in the N and S regions shown in  Fig.~\ref{fig21}.  
At $\beta=7$ the DOS in the N layer is nearly equal to that of of normal state but  
${\rm Re} F$ does not vanish. Due to the contribution of ${\rm Re} F$ in the spectral function $M$ in Eq. (\ref{M}), 
the N layer still has proximity induced pair correlations but $R_s$ becomes larger than $R_s$ of a fully decoupled N layer with $\beta\to\infty$. 
As the field increases, ${\rm Re} F$ decreases, so $R_s$ approaches $R_s$ of the normal state.  

Shown in Fig.~\ref{fig22} is the effect of diffusivity ratio $D_n/D_s$ on $R_s(H_0)$ calculated for different values of $\beta$. 
As the ratio $D_n/D_s$ increases, the minimum first shifts to lower fields because the effect of current pair-breaking in the N layer becomes more pronounced so that 
$R_s(H_0)$ turns to increase at lower fields. 
However, this trend reverses at larger $\beta$ as it is evident from Fig.~\ref{fig22}(c), where the minimum disappears and the field dependence of $R_s(H_0)$ is determined by the behavior of the Green functions discussed above in relation to $R_s(H_0)$ for a nearly decoupled N layer 
shown in Figs.~\ref{fig20}(b) and Fig.~\ref{fig21}. 

\begin{figure}[tb]
   \begin{center}
   \includegraphics[width=0.95\linewidth]{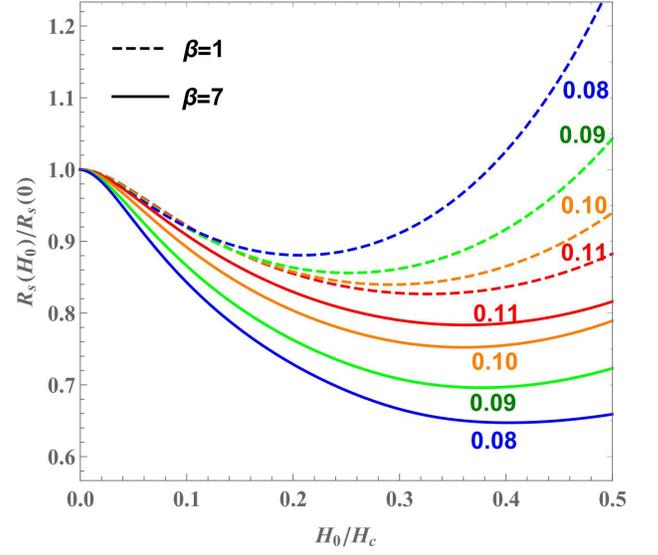}
   \end{center}\vspace{0cm}
   \caption{
$R_s(H_0)$ calculated for $T /\Delta= 0.08, 0.09, 0.10, 0.11$, 
$\alpha=0.001$, $\beta=1, 7$, $\Gamma_{\rm N}/\Delta= \Gamma_{\rm S}/\Delta = 0.005$, 
$\Omega/\Delta = 11$, $D_n = 5 D_s$, $\lambda=10\xi_s$, 
and $\omega /\Delta = 0.001$. 
   }\label{fig23}
\end{figure}
\begin{figure}[tb]
   \begin{center}
   \includegraphics[width=0.95\linewidth]{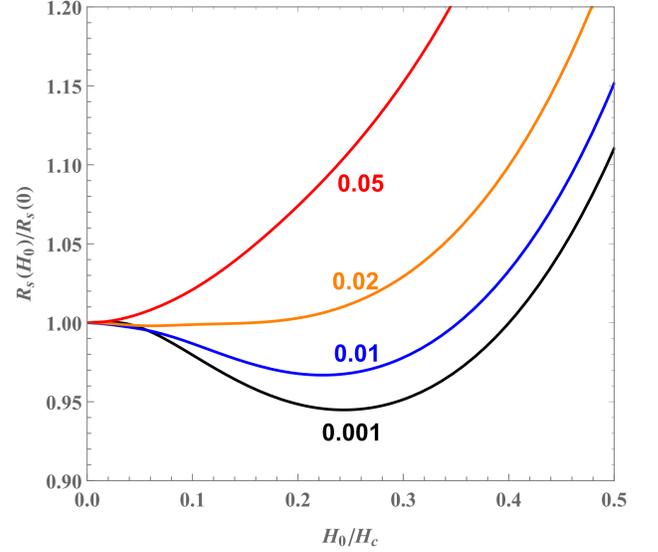}
   \end{center}\vspace{0cm}
   \caption{
$R_s(H_0)$ calculated for $\omega/\Delta = 0.001, 0.01, 0.02, 0.05$, 
$\alpha=0.05$, $\beta=0.5$, $\Gamma_{\rm N}/\Delta= \Gamma_{\rm S}/\Delta = 0.005$, 
$\Omega/\Delta = 11$, $D_n = 0.5 D_s$, $\lambda=10\xi_s$, 
and $T /\Delta= 0.1$. 
   }\label{fig24}
\end{figure}

Now we turn to the effects of temperature and the rf frequency on the field dependence of $R_s(H_0)$. 
Fig.~\ref{fig23} shows that the evolution of $R_s(H_0)$ with $T$ can 
change as the interface resistance increases and the N layer becomes more decoupled. For instance, at $\beta=1$,  the minimum in $R_s(H_0)$ 
becomes shallower and shifts to lower fields as $T$ decreases. This behavior of $R_s(H_0,T)$ is controlled by the same mechanism discussed above in relation to 
Figs.~\ref{fig7} and \ref{fig12}. 
However, this evolution of $R_s(H_0)$ with $T$ reverses for $\beta=7$ corresponding to the solid curves in Fig.~\ref{fig23}. 
In this case the minimum becomes more pronounced and shifts to higher fields as $T$ decreases. 
Here the field dependence of $R_s(H_0)$ is mostly determined by the effect of the N layer similar to that is shown in Fig.~\ref{fig22}(c), whereas the contribution of the S region diminishes at $T$ decreases. 
The effect of frequency on $R_s(H_0)$ shown in Fig.~\ref{fig24} is qualitatively similar to that was discussed above in relation to Figs.~\ref{fig8} and \ref{fig13}. In both cases our calculations do not take into account nonequilibrium effects which become more pronounced at lower temperatures and higher frequencies.

\section{Discussion}

In this work we address mechanisms of nonlinear RF losses and show that $R_s(H_0)$ can be reduced by optimizing the DOS using a multitude of pairbreaking mechanisms, including the N-S or S-I-S$^\prime$ surface nanostructuring, impurity management or the rf Meissner currents. Here the nonmonotonic field dependence of $R_s(H_0)$ results from the broadening of the DOS gap peaks by the pair-breaking rf currents ~\cite{2014_Gurevich_PRL}. Combining the rf current pairbreaking with pairbreaking caused by the material features~\cite{2017_Gurevich_SUST, 2017_Gurevich_Kubo} opens up opportunities to tune the field dependence of $R_s(H_0)$ by materials processing and surface treatment. For instance, magnetic impurities shift the minimum in $R_s(H_0)$ to lower fields and can reduce $R_s(H_0)$ in a wide range of $H_0$ (Figs.~\ref{fig5}). Bulk subgap states can give rise to a residual surface resistance while reducing $R_s$ at higher temperatures (see Figs.~\ref{fig10} and \ref{fig11}). 
A S-I-S$^\prime$ multilayer with the optimal thickness $d\sim \lambda'$ can shift the minimum in $R_s(H_0)$ to higher fields exceeding the superheating field of the S-substrate. A proximity-coupled N layer not only affects the field position and the depth of the minimum in $R_s(H_0)$ but can also reduce the surface resistance below $R_s$ for an ideal  surface without the N layer (see Figs.~\ref{fig19}, \ref{fig20}, and \ref{fig22}). 

These results can explain why the field dependence of the quality factor $Q(H_0) \propto 1/R_s(H_0)$ can be very sensitive to the materials processing. 
For instance, magnetic impurities in Nb have been attributed to oxygen vacancies in ${\rm Nb_2 O_5}$, while other impurities can segregate at the surface during chemical polishing and electro polishing ~\cite{2005_Casalbuoni, 2011_Proslier}. It has been found that low-temperature baking increases the Curie constant, which in turn increases the spin flip parameter $\Gamma_p$ and affects $R_s(H_0)$ curves  as shown in Fig.~\ref{fig5}. 
Tunneling experiments have shown that the DOS is strongly affected by infusion of Ti at high-temperatures,
which sharpens the DOS peaks and results in a pronounced minimum in $R_s(H_0)$~\cite{2013_Dhakal}. 
These observations are consistent with the results of this work in which the behavior of $R_s(H_0)$ is controlled by the DOS at the surface and the values of $\Gamma$ and $\Gamma_p$ which manifest themselves in different $R_s(H_0)$ curves shown in Fig.~\ref{fig10}. 

A common feature of many superconducting materials is a thin proximity-coupled N layer on the surface, for example, metallic suboxides underneath the dielectric ${\rm Nb_2 O_5}$ pentooxide layer on the Nb surface. Chemical or heat treatments change the properties of the N layer, particularly the thickness, impurity concentration, and the interface contact resistance~\cite{1992_Robertazzi, 1993_Ekin}. As a result, the materials processing used for superconducting resonator cavities ~\cite{2014_Padamsee_review, 2012_Aontoine_review}, 
such as electropolishing~\cite{1989_Saito} and heat treatments at high temperatures~\cite{2013_Dhakal, 2013_Grassellino} and low temperatures~\cite{2017_Grassellino, 2018_Dhakal, 2018_Grassellino}, 
can significantly affect $R_s(H_0)$. These observations are qualitatively consistent with our calculations which show that various $R_s(H_0)$ curves similar to those observed on the Nb cavities can occur, depending on the N layer properties and the value of $R_B$, as illustrated by Figs.~\ref{fig19}, \ref{fig20}, and \ref{fig22}. Yet a quantitative comparison of our theory with experiment would require independent measurements of multiple parameters characterizing a particular material, for example, the thickness and conductivity of the N layer, $R_B$, $\Gamma_N$ and $\Gamma_S$, as well as the way these parameters change after different materials treatments. This information can be extracted from different surface characterization techniques, for example, scanning tunneling microscopy or X-ray photoelectron spectroscopy.   Such surface measurements combined with the rf measurements of $Q(H_0)$ and the theory analysis could clarify whether it is a reduction of the thickness of the N layer or contact resistance or the Dynes parameters in the bulk which plays a major role in the development of the field-induced reduction of $R_s(H_0)$ after the Ti or N infusion or other heat treatments in Nb.  

Tuning of $R_s(H_0)$ can also be done by depositing a S$^\prime$ layer with higher-$H_c$ and $\lambda'>\lambda$ on the S surface, which can withstand magnetic fields of the order of the superheating field $H_s^\prime$ while significantly increasing the field onset of vortex penetration \cite{mlag1}. Moreover, the current counterflow caused by the S substrate in the S$^\prime$ layer with 
$\lambda'>\lambda$ allows the S-I-S$^\prime$ structure to remain in the Meissner state at $H_0$ exceeding both superheating fields $H_s$ and $H_s^\prime$ at an optimum thickness $d\sim \lambda'$~ \cite{mlag2,mltk1,mltk2}. This counterflow effect can shift the minimum in $R_s(H_0)$ to higher fields exceeding $H_s$ of the S-region at $d\simeq \lambda'$, as shown in Fig.  \ref{fig16} for a Nb$_3$Sn-I-Nb structure. 
Also, the Nb covered with a dirtier Nb layer, which models the Nb surface after low-temperature baking ~\cite{muonSR}, mitigates the high-field increase of $R_s (H_0)$ ~\cite{2017_Gurevich_SUST,hs3} as shown in Fig.~\ref{fig15}. 
This conclusion is also qualitatively consistent with experiments on the Nb cavities  (see, e.g., reviews \onlinecite{2012_Aontoine_review,2014_Padamsee_review} or Refs. \onlinecite{2018_CEH, 2018_Grassellino} for recent developments).  

For a superconductor with an ideal surface, the minimum in $R_s(H_0)$ shown in Figs.~\ref{fig7} and \ref{fig12} becomes shallower as $T$ decreases. 
On the other hand, for a weakly-coupled N layer with a large contact resistance $(\beta \gtrsim 1)$, the minimum in $R_s(H_0)$ becomes more pronounced as  the temperature decreases (see Fig. \ref{fig23}) because the N layer contribution to $R_s(H_0)$ becomes dominant at low $T$. 
These conclusions follow from our calculations in which the equilibrium distribution function of quasiparticles $f_0(\epsilon)$ was assumed. 
As shown in Sec. II C, this condition is satisfied at low frequencies $\hbar\omega \ll k_BT$, the subgap states and minigaps in the N layer
accelerate the energy relaxation of quasiparticles reducing the deviation of $f(\epsilon)$ from the Fermi distribution.  Yet the calculations of $R_s(H_0)$ at low temperatures and higher frequencies would require the non-equilibrium quasiparticle distribution function obtained by solving the kinetic equations for strong RF fields \cite{kopnin,kramer,lo}. 

We considered different ways of minimization of $R_s(H_0)$ by optimizing the DOS at the surface. Yet 
it should be noted that the optimum parameters actually depend on the rf amplitude. 
For instance, Fig.~\ref{fig6} shows that the optimum spin flip parameter $\Gamma_p$ is different for different $H_0$, 
although the optimum concentration of paramagnetic impurities is roughly determined by the spin-flip scattering mean free path $\ell_s \sim 10^2 \xi_0$, at which
$R_s$ can be significantly reduced in a wide range of $H_0$. The subgap states affect $R_s$  in a way similar to that of paramagnetic impurities at higher temperatures. 
While the physics and materials mechanisms behind the Dynes parameter $\Gamma$ are not well understood, 
tunneling spectroscopy can give valuable insights into how  $\Gamma$ is affected by various materials processing. The observed values of $\Gamma$ can then be used as input parameters in the theory of this work to identify the optimum materials treatments which minimize the nonlinear surface resistance. 

\appendix

\section{Pair potential} \label{appendix:Delta}

For a superconductor with an ideal surface, the solution of the thermodynamic Usadel equation (\ref{eq:usadel}) 
in the first order in $s/\Delta\ll 1$ is given by:
\begin{equation}
\theta = \sin^{-1} \frac{\Delta}{\sqrt{\tilde{\omega}_n^2 + \Delta^2}}-\frac{s\Delta\tilde{\omega}_n}{(\Delta^2+\tilde{\omega}_n^2)^{3/2}}.
\label{thets} \\
\end{equation}
Substituting Eq. (\ref{thets}) into Eq. (\ref{eq:self-consistency}) yields the equation for $\Delta$ in the case of weak pairbreaking:
\begin{gather}
\frac{1}{g}=2\pi T\sum_{\omega_n>0}^\Omega\frac{1}{\sqrt{(\omega_n + \Gamma)^2 + \Delta^2}}-
\nonumber \\
-2\pi T\sum_{\omega_n>0}\frac{s(\omega_n+\Gamma)^2}{[\Delta^2+(\omega_n+\Gamma)^2]^2}.
\label{delgs}
\end{gather}
At $T\ll T_c$ we replace the summation in Eq. (\ref{delgs}) with integration over $\omega_n$, express $1/g=\ln(2\Omega/\Delta_0)$ in terms of the gap 
$\Delta_0$ at $s=T=0$ and obtain:
\begin{gather}
\ln\frac{\Delta_0}{\Gamma+\sqrt{\Gamma^2+\Delta^2}}=
\nonumber \\
\frac{s}{4\Delta}\left[\pi+\frac{2\Gamma\Delta}{\Gamma^2+\Delta^2}-2\tan^{-1}\frac{\Gamma}{\Delta}\right],\quad T=0.
\label{del0}
\end{gather}
At $s=0$ Eq. (\ref{del0}) yields:
\begin{equation}
\Delta^2=\Delta_0(\Delta_0-2\Gamma).
\label{del00}
\end{equation} 
Hence, the subgap states reduce the superfluid density and fully suppress superconductivity at $\Gamma>\Delta_0/2$. For weak DOS broadening and 
current pairbreaking considered in this work, the solution of Eq. (\ref{del0}) in the first order in $s$ and $\Gamma$ is given by:
\begin{equation}
\Delta=\Delta_0-\Gamma-\frac{\pi s}{4}.
\label{delt}
\end{equation} 


\section{Nonlinear conductivity} \label{appendix:kineq}
Here we present the derivation of the time-averaged $\sigma_1(H)$ following the approach of Ref. (\onlinecite{2014_Gurevich_PRL}) based on  
the time-dependent Usadel equations for the $4\times 4$ quasiclassical Greens function $\check{G}(\textbf{r},t,t')$ \cite{lo,kopnin,1999_Belzig_review}:
\begin{gather}
\partial _{t}\check{\sigma}_z\check{G}+\partial _{t'}\check{G}\hat{\sigma}_z=D\check{\Pi}\circ (\check{G}\circ 
\check{\Pi}\check{G})-[\check{\Delta}%
,\check{G}],  \label{usad} \\
\check{G}=\left( 
\begin{array}{cc}
\hat{G}^{R} & \hat{G}^{K} \\ 
0 & \hat{G}^{A}
\end{array}
\right) ,\qquad 
\check{\Delta}=\left( 
\begin{array}{cc}
 \hat{\Delta} & 0 \\
0  & \hat{\Delta}
\end{array}
\right), 
\label{def} \\
\hat{G}^{R,A}=\left( 
\begin{array}{cc}
G^{R,A} & F^{R,A} \\ 
F^{\dagger R,A} & -G^{R,A}
\end{array}
\right), \quad 
\hat{\Delta}=\left( 
\begin{array}{cc}
0 & \Delta \\
\Delta^* & 0
\end{array}
\right),
\label{gf}
\end{gather}
where $\hat{G}^{R}$ and $\hat{G}^{A}$ are the retarded and advanced Green functions, $\check{G}\circ \check{G}=\check{1}$,
$\hat{G}^{K}=\hat{G}^{R}\circ\hat{f}-\hat{f}\circ \hat{G}^{R}$ is the Keldysh function expressed in terms of a distribution 
function of quasiparticles, $\hat{f}(\epsilon,t)$, the hat denotes matrices in the Nambu space, 
$\hat{\Pi}=\nabla +i\pi \mathbf{A}\hat{\sigma}_z/\phi _{0}$,  
 $\check{\sigma}_z$ is a diagonal matrix composed of the Pauli matrices $\hat{\sigma}_z$, and $\circ$ means time convolution. 
For a dirty type-II superconductor with $\lambda\gg\xi$,  the relation between the
current density and the vector potential is local but time-dispersive:
\begin{gather}
{\bf J}({\bf r},t)=\frac{\sigma_n}{2c}\mbox{Im}\mbox{Tr}\int D(t,t',\mathbf{r}){\bf A}({\bf r},t')dt', 
\label{a3}\\
D=\int\{\hat{G}_{z}^{R}(t,t^{\prime })\bigl[\hat{G}_{z}^{R}(t^{\prime },t_{1})f(t_{1},t) 
-\hat{G}_{z}^{A}(t_{1},t)f(t^{\prime },t_{1})\bigr] \nonumber \\
+\bigl[f(t_{1},t^{\prime})\hat{G}_{z}^{R}(t,t_{1})
-f(t,t_{1})\hat{G}_{z}^{A}(t_{1},t^{\prime })\bigr]\hat{G}_{z}^{A}(t^{\prime },t)\}dt_1,
\label{a4}
\end{gather}
where $\hat{G}_z=\hat{\sigma}_z\cdot \hat{G}$. In Eq. (\ref{a4}) the gradient terms $\nabla G$ are neglected, and the matrix $\hat{f}$ reduces to a single distribution function since the transverse electromagnetic field does not cause the electron-hole imbalance.  

We use the standard mixed Fourier representation 
\begin{equation}
G(t,t^{\prime})=\int_{-\infty}^{\infty}G\left(\epsilon, t_{0}\right)
e^{i\epsilon(t^{\prime}-t)}\frac{d\epsilon}{2\pi},
\end{equation}
where $t_0=(t+t')/2$. Expanding the time convolution in small derivatives with respect to $t'$ in Eq. (\ref{a4}) for slow variations of $\textbf{A}(\textbf{r},t')$ and 
neglecting linear in $\omega$ terms yields:
\begin{gather}
D(t,t^{\prime})  =\mbox{Tr}\int  f(\epsilon,t_{0})\big[ e^{i(\epsilon-\epsilon^{\prime}%
)(t-t^{\prime})}\hat{G}_{z}^{R}(\epsilon^{\prime},t_{0}) +
\nonumber \\
e^{i(\epsilon
^{\prime}-\epsilon)(t-t^{\prime})}\hat{G}_{z}^{A}(\epsilon^{\prime}%
,t_{0})\big]
 \big[\hat{G}_{z}^{R}(\epsilon,t_{0})-\hat{G}_{z}^{A}(\epsilon
,t_{0})\big] \frac{d\epsilon d\epsilon^{\prime}}%
{(2\pi)^{2}}%
\end{gather}

The time-averaged nonlinear conductivity $\sigma_1=\bar{\sigma}$ is defined in terms of the mean dissipated power 
$q=\bar{\sigma}E_0^2/2$ induced by the ac electric field $\mathbf{E}=-c^{-1}\partial_{t}\mathbf{A=E}_{0}\sin\omega t$ for which
$\mathbf{A}(t)=(c\mathbf{E}_{0}/\omega)\cos\omega t$. Here
\begin{gather}
q=\lim_{t_{m}\rightarrow\infty}\frac{1}{2t_{m}}\int_{-t_{m}}^{t_{m}}%
\mathbf{J}(t)\mathbf{E}(t)dt= 
 \nonumber \\
\frac{i\sigma_{n}E_{0}^{2}}{4t_{m}\omega}%
\int_{-t_{m}}^{t_{m}}\sin\omega tdt\int dt^{\prime}%
\cos\omega t^{\prime}\int\frac{d\epsilon d\epsilon^{\prime}}{(2\pi)^{2}}
\nonumber \\
\!\!\!\left[ e^{i(\epsilon-\epsilon^{\prime})(t-t^{\prime})}P_{1}(\epsilon
,\epsilon^{\prime},t_{0})+e^{i(\epsilon^{\prime}-\epsilon)(t-t^{\prime})}%
P_{2}(\epsilon,\epsilon^{\prime},t_{0})\right]\!,
\label{a7}
\end{gather}
where 
\begin{gather}
P_{1}=\mbox{Tr}\hat{G}_{z}^{R}
(\epsilon^{\prime},t_{0})\big[\hat{G}_{z}^{R}(\epsilon,t_{0})-\hat{G}%
_{z}^{A}(\epsilon,t_{0})\big]f(\epsilon,t_{0}),
\label{P1} \\
P_{2}=\mbox{Tr}\hat{G}_{z}^{A}(\epsilon^{\prime},t_{0})\big[
\hat{G}_{z}^{R}(\epsilon,t_{0})-\hat{G}_{z}^{A}(\epsilon,t_{0})\big]
f(\epsilon,t_{0}).
\label{P2}
\end{gather}
Changing variables $t=t_{0}+t_{1}/2$ and $t^{\prime
}=t_{0}-t_{1}/2,$ yields 
\begin{gather}
q=\lim_{t_{m}\rightarrow\infty}\frac{i\sigma_{n}E_{0}^{2}}{8t_{m}\omega}%
\int_{-t_{m}}^{t_{m}} dt_{0}dt_{1}(\sin\omega t_{1}+\sin2\omega t_{0}) \cdot \nonumber \\
\int\frac{d\epsilon
d\epsilon^{\prime}}{(2\pi)^{2}}\left[  e^{i(\epsilon-\epsilon^{\prime})t_{1}
}P_{1}(\epsilon,\epsilon^{\prime},t_{0})+e^{i(\epsilon^{\prime}-\epsilon
)t_{1}}P_{2}(\epsilon,\epsilon^{\prime},t_{0})\right] 
\end{gather}
In the dirty limit $P_{1}(\epsilon,\epsilon^{\prime},t_{0})$ and
$P_{2}(\epsilon,\epsilon^{\prime},t_{0})$ are even functions of $t_{0}$, so
only $\sin\omega t_{1}$ contributes:
\begin{gather}
\sigma_1  =
\frac{\sigma_{n}}{4\pi}\int_{0}^{\pi/\omega}dt\int\frac{d\epsilon}{2\pi}
\big[  P_{1}(\epsilon,\epsilon+\omega,t)-
\nonumber \\
P_{1}(\epsilon+\omega
,\epsilon,t)+P_{2}(\epsilon+\omega,\epsilon,t)-P_{2}(\epsilon,\epsilon
+\omega,t)\big].
\label{sig}
\end{gather}
Using $\mbox{Tr}\hat{G}_{z}\cdot \hat{G}_{z}=2G\cdot G+F\cdot F^{\dag}+F^{\dag}\cdot F$ gives
\begin{gather}
P_{1}(\epsilon,\epsilon^{\prime},t)-P_{2}(\epsilon,\epsilon^{\prime},t
)=2f(\epsilon,t)\{[G^{R}_{\epsilon^{\prime}}-G^{A}_{\epsilon^{\prime}}]\cdot
\nonumber \\
[G^{R}_{\epsilon}-G^{A}_{\epsilon}]
+[F^{R}_{\epsilon^{\prime}}-F^{A}_{\epsilon^{\prime}
}][F^{R}_{\epsilon}-F^{A}_{\epsilon}]\},
\label{pp}
\end{gather}
Because $\mbox{Re}(G,F)^{R}=-\mbox{Re}(G,F)^A$, and $\mbox{Im}(G,F)^{R}=\mbox{Im}(G,F)^A$, we can present Eqs. (\ref{sig}) and (\ref{pp}) in the form
\begin{gather}
\sigma_1=\frac{\sigma_n}{\pi}\int_0^{\pi/\omega}\int_{-\infty}^{\infty}d\epsilon [f(\epsilon,s)-f(\epsilon+\omega,s)]\cdot
\nonumber \\
[\mbox{Re}G^R(\epsilon+\omega)\mbox{Re}G^R(\epsilon)+\mbox{Re}F^R(\epsilon+\omega)\mbox{Re}F^R(\epsilon)].
\label{sigmat}
\end{gather}
In the parameterization $G^{R,A}_{\epsilon}=\pm\cosh u\cos v+i\sinh u\sin
v$ and $F^{R,A}_{\epsilon}=\pm\sinh u\cos v+i\cosh u\sin v$, Eq. (\ref{sigmat}) becomes
\begin{gather}
\sigma_1=\frac{\sigma_{n}}{\pi}\int_{0}^{\pi/\omega}dt\int_{-\infty}^{\infty}d\epsilon [f(\epsilon,s)-f(\epsilon+\omega,s)]\cdot
\nonumber \\
\cosh (u_{\epsilon}+ u_{\epsilon+\omega})\cos v_{\epsilon}\cos v_{\epsilon +\omega}.
\label{sigt}
\end{gather}
If $s=0$, we have $\cos v_{\epsilon}=1$, $\cosh
u_{\epsilon}=$ $\epsilon/\sqrt{\epsilon^{2}-\Delta^{2}}$, $\sinh u_{\epsilon
}=\Delta/\sqrt{\epsilon^{2}-\Delta^{2}}$ at $|\epsilon|>\Delta$. If $|\epsilon|<\Delta$ the integrand  in Eqs. (\ref{sigmat}) and (\ref{sigt}) vanishes and
$\bar{\sigma}$ reproduces the Mattis-Bardeen result for $\omega<\Delta$:
\begin{equation}
\sigma_1=\frac{2\sigma_{n}}{\omega}
\int_{\Delta}^{\infty}\frac{[\epsilon(\epsilon+\omega)+\Delta^{2}
][f(\epsilon)-f(\epsilon+\omega)]d\epsilon}{\sqrt{\epsilon^{2}-\Delta^{2}
}\sqrt{(\epsilon+\omega)^{2}-\Delta^{2}}}.
\end{equation}
At $\exp(-\Delta/T)\ll1$ the main
contribution to this integral comes from a narrow range 
$\epsilon-\Delta\sim T\ll\Delta$, where one can take $z=\epsilon-\Delta$ and
$\epsilon^{2}-\Delta^{2}\approx 2\Delta z$. Then
\begin{equation}
\sigma_1=\frac{2\sigma_{n}}{\omega}\left(  1-e^{-\omega/T}\right)
e^{-\Delta/T}\int_{0}^{\infty}\frac{e^{-z/T}dz}{\sqrt{z(z+\omega)}}.
\end{equation}
Hence  
\begin{equation}
\sigma_1=\frac{4\sigma_{n}\Delta}{\omega}\sinh\left[  \frac
{\omega}{2T}\right]  K_{0}\left[  \frac{\omega}{2T}\right]
e^{-\Delta/T}.
\label{mb}
\end{equation}
Here a modified Bessel function reduces to 
$K_0(\omega/2T) \simeq \ln(4T/\omega) - \gamma_E$ at $\omega \ll 2T$,
where $\gamma_E = 0.577$ is the Euler constant. Then Eq. (\ref{mb}) gives $R_s=\mu_0\omega^2\lambda^3\sigma_1/2$ which reproduces Eq. (\ref{eq:Rs_MB}) with
$C = 4e^{-\gamma_E}\approx 9/2$.


\section{Proximity coupled normal layer} \label{appendix:Proxi}

To calculate $\theta$ in the N region,  we solve the Usadel equation with $\Delta=0$: 
\begin{equation}
\theta'' = s\sin\theta\cos\theta + r \tilde{\omega}_{{\rm N}n}\sin\theta.
\label{c1} 
\end{equation}
Here $r=D_{\rm S}/D_{\rm N}$ and $\tilde{\omega}_{{\rm N}n} = \omega_n + \Gamma_{\rm N}$, and we use the dimensionless form with 
dimensionless lengths $x\to x/\xi_s$ and energies $\omega_n\to\omega_n/\Delta$, $s\to s/\Delta$, $\Gamma\to \Gamma/\Delta$. 

If $d \ll \xi_{n}$ the function $\theta(x)$ varies weakly across the N layer, so 
the solution of Eq. (\ref{c1}) satisfying the boundary condition $\theta'(-d)=0$ can be approximated by
\begin{equation}
\!\!\!\theta(x) = \theta_{-} + \frac{x(x+2d)}{2} \bigl( s\sin\theta_{-} \cos\theta_{-} + r\tilde{\omega}_{{\rm N}n} \sin\theta_{-} \bigr) . 
\label{c2}
\end{equation}
The boundary condition~(\ref{eq:BC1}) at $x=0$ takes the form
\begin{gather}
\sin(\theta_0 -\theta_{-}) = \beta \eta, 
\label{eq:BC1-2} \\
\eta = \frac{s}{r} \sin\theta_{-} \cos\theta_{-} + (\omega_n + \Gamma_{\rm N}) \sin\theta_{-},
\label{eta}
\end{gather}
where $\alpha$ and $\beta$ are defined by Eqs.~(\ref{eq:alpha}) and (\ref{eq:beta}).

To obtain the relation between $\theta_-$ at the N side of the boundary and $\theta_0'$ at the S side of the boundary, 
we solve the Usadel equation in the S region ($x \ge 0$):
\begin{equation}
\theta'' = s\sin\theta\cos\theta + \tilde{\omega}_{{\rm S}n}\sin\theta -\Delta(x) \cos\theta . 
\label{eq:T_Usadel_S}
\end{equation}
Here $\tilde{\omega}_{{\rm S}n} = \omega_n + \Gamma_{\rm S}$. In the vicinity of the NS interface, 
the proximity effect causes a weak short-range disturbance $\Delta(x)=\Delta_s + \delta\Delta(x)$. 
we can approximate by~\cite{2017_Gurevich_Kubo}:
\begin{eqnarray}
\delta \Delta = - \Psi \delta(x)
\label{ddel}
\end{eqnarray}
with $\Psi$ to be determined self-consistently. 
Integrating Eq.~(\ref{eq:T_Usadel_S}) from $0$ to $+0$ and using $\theta'(-0)=\alpha\eta$, 
we obtain the boundary condition for Eq. (\ref{eq:T_Usadel_S}) at $x=+0$:
\begin{eqnarray}
\theta'(+0) = \alpha \eta + \Psi \cos\theta_0 .
 \label{eq:BC2-3}
\end{eqnarray}
A thin N layer produces a weak disturbance of the pair potential, $\Psi \ll 1$ ~\cite{2017_Gurevich_Kubo}, so 
Eq. (\ref{eq:T_Usadel_S})  can be linearized in $\delta \theta(x) = \theta(x) - \theta_s$, giving $\delta \theta'' = -k^2 \delta \theta$ where
$k$ is defined by Eq. (\ref{k2}). Hence,  
\begin{equation}
\delta \theta (x) = \delta \theta_0 e^{-kx},\qquad 
\delta \theta_0 = - \frac{1}{k} ( \alpha \eta + \Psi \cos\theta_0 ) 
\label{eq:dTheta0}
\end{equation}
Substituting  Eqs. (\ref{ddel}) and (\ref{eq:T_Usadel_S}) into Eq. (\ref{eq:self-consistency}) linearized with respect to $\delta\theta_0$ and integrating it over $x$ yields the self-consistency 
Eq. (\ref{eq:Psi}) for $\Psi$. 

Consider now the retarded Green's functions.  
The solution of the real frequency Usadel equation in the N region is similar to that of Eq. (\ref{c2}):
\begin{eqnarray}
\theta(x) = \theta_{-} + \frac{x(x+2d)}{2} \bigl( s\sinh\theta_{-} \cosh\theta_{-}  -i r\tilde{\epsilon}_{\rm N} \sinh\theta_{-} \bigr) , 
\label{eq:R_Usadel_solN} \nonumber \\
\end{eqnarray}
where $\tilde{\epsilon}_{\rm N} = \epsilon + i\Gamma_{\rm N}$.  In the S region we have: 
\begin{equation}
\theta'' = s \sinh\theta \cosh\theta  -i \tilde{\epsilon}_{\rm S} \sinh\theta +i [\Delta_s -\Psi \delta(x) ] \cosh\theta, 
\label{uspsi}
\end{equation}
where $\tilde{\epsilon}_{\rm S} = \epsilon + i\Gamma_{\rm S}$, and $\Psi$ is given by Eq.~(\ref{eq:Psi}). 
Substituting $\theta=u+iv$ into Eq. (\ref{uspsi}) and separating real and imaginary parts at $x=+0$ yields Eqs. (\ref{x-dependence1}) and (\ref{x-dependence2}).

The equation connecting the boundary values $\theta_-$ and $\theta_0$ readily follows from Eqs. (\ref{eq:BC1-2}) and (\ref{eta}): 
\begin{gather}
\beta \biggl( \frac{s}{r}\sinh\theta_{-}\cosh\theta_{-} - i \tilde{\epsilon} \sinh\theta_{-} \biggr) \nonumber \\
= \sinh\theta_0 \cosh\theta_{-}  - \cosh\theta_0 \sinh\theta_{-} , 
\label{thet00}
\end{gather}
Taking here $\theta_0 = u_0 + iv_0$ and $\theta_{-}= u_{-}+iv_{-}$ and separating real and imaginary parts yields Eqs. (\ref{bc1re}) and (\ref{bc1im}).
Another boundary condition is obtained by integrating Eq. (\ref{uspsi}) from $x=-0$ to $x=+0$, which yields
$r\theta'_--\theta_0'=i\Psi\cosh\theta_0$. Substituting here $\theta'_-=\alpha(s\sinh\theta_-\cosh\theta_--ir\tilde{\epsilon}_n\sinh\theta_-)$ from Eq. (\ref{eq:R_Usadel_solN}) gives
\begin{equation}
\theta_0'=\alpha\left[\frac{s}{2r}\sinh 2\theta_--i\tilde{\epsilon}_N\sinh\theta_- \right]-i\Psi\cosh\theta_0
\end{equation}
Separating here real and imaginary parts results in Eqs. (\ref{bc2re}) and (\ref{bc2im}).

\begin{acknowledgments}

The work of T. K. was supported by Japan Society for the Promotion of Science (JSPS) KAKENHI Grant Number JP17H04839 and JP17KK0100. 
The work of A. G. was supported by NSF under Grant No. PHY-1416051 and DOE under Grant No DE-SC100387-020. 

\end{acknowledgments}


%

\end{document}